\documentclass[usenatbib]{mnras}
\usepackage{graphicx}
\usepackage{float}
\usepackage{longtable}
\usepackage{threeparttable}
\usepackage{amsmath}
\usepackage{color}
\usepackage{subcaption}
\usepackage{afterpage}
\usepackage{enumitem}
\usepackage{pifont}

\newcommand{\Nc}{N$_{\mathrm{C}}$}
\newcommand{\AvgNc}{$\langle\mathrm{N_{C}}\rangle$}

\defcitealias{Maragkoudakis2020}{MPR20}

\title[PAH spectral variations]{Spectral variations among different scenarios of PAH processing or formation}
\author[A. Maragkoudakis et al.]
{A. Maragkoudakis$^{1,2,3}$\thanks{E-mail: maragkoudakis.alex@gmail.com}, E. Peeters$^{3,4,5}$ and A. Ricca$^{1,5}$\\
    $^{1}$NASA Ames Research Center, MS 245-6, Moffett Field, CA 94035-1000, USA\\
    $^{2}$Oak Ridge Associated Universities, Oak Ridge, TN, USA\\
	$^{3}$Department of Physics and Astronomy, University of Western Ontario, London, ON, N6A 3K7, Canada\\
	$^{4}$Centre for Planetary Science and Exploration, University of Western Ontario, London, Ontario N6A 5B7, Canada \\
	$^{5}$Carl Sagan Center, SETI Institute, 189 Bernardo Ave., Mountain View, CA 94043, USA}

\begin{document}
	
	\date{}
	
	\maketitle

	\begin{abstract}
		We examine the variations in the spectral characteristics and intensities of PAHs in two different scenarios of PAH processing (or formation): \textit{(1)} small PAHs are being destroyed (or equivalently large PAHs are being formed, referred to as SPR i.e. small PAHs removed), and \textit{(2)} large PAHs are being destroyed (or equivalently small PAHs are being formed referred to as LPR i.e. large PAHs removed). PAH emission was measured considering both the presence or absence of plateau components. The variation in the PAH band intensities as a function of the average number of carbon atoms \AvgNc{} has the highest dynamic range in the SPR case suggesting that smaller PAHs have higher impact on the PAH band strengths. The plateaus show overall declining emission with \AvgNc{}, and their higher dynamic range in the SPR case also suggests that smaller PAHs are mainly contributing to the plateau emission. The 7.7/(11.0+11.2) \micron{} PAH band ratio presents the least amount of variance with the lowest dynamic range, rendering this ratio as the better choice for tracing PAH charge. The 3.3/(11.2+11.0) \micron{} PAH band ratio is the only ratio that has both a monotonic variance and fully separated values among the SPR and LPR scenarios, highlighting its efficiency as PAH size tracer but also allowing the characterization of the dominant scenario of processing or formation in a given region or source. We present new PAH charge -- size diagnostic diagrams, which can provide insights on the average, maximum, or minimum \Nc{} within astrophysical sources.

	\end{abstract}
	 
	
	\section{Introduction}
	
Emission in the so-called aromatic infrared bands (AIBs) often dominates the mid-infrared ($ 3 - 20\,\micron$) spectra of various astrophysical sources \citep[e.g.][]{Hony2001, Peeters2002, Smith07b, Gordon2008}. These prominent emission bands at 3.3, 6.2, 7.7, 8.6, 11.2, 12.7, 16.4, and 17.4 $ \micron $, with weaker features at surrounding wavelengths, are generally attributed to the family of Polycyclic Aromatic Hydrocarbons (PAHs) and are due to vibrational emission of the carbon-carbon (C-C) and carbon-hydrogen (C-H) bonds in the PAH molecules upon the absorption of a FUV photon \citep{Leger1984, Allamandola1985}. Although alternative carriers have been proposed for these emission features (e.g. hydrogenated amorphous carbons--HAC, \citealt{Duley1985, Jones2013}; quenched carbonaceous composite, \citealt{Sakata1984}), the first detection of a few small PAHs in the molecular cloud TMC-1 \citep[][]{McGuire2018,McGuire2021,McCarthy2021, Cernicharo2021} as well as the confirmed detection of the C$ _{60} $ and C$ _{70} $ fullerene emission \citep{Cami2010, Sellgren2010} in astrophysical spectra provide strong support for the presence of PAH molecules in space.

The emission from PAH molecules can offer unique insights in the study of the local physical conditions of their residing environments \citep[e.g.][]{Galliano2008b, Boersma2015, Pilleri2015, Stock2016, Stock2017, Zang2022, Maragkoudakis2022}. For instance, the intensity of the UV radiation field ($ G_{0} $; given in units of the Habing field: $1.2 \times 10^{-4}$ erg cm$^{-2}$ s$^{-1}$ sr$^{-1}$) can be assessed through the empirical calibration of the $I_{6.2}/I_{11.2}$ PAH band ratio and the ionization parameter $\gamma$  \citep[$G_{0} T^{1/2}/n_{e}$ where $T$ is the gas temperature and $n_{e}$ the electron density; e.g.][]{Galliano2008b, Maragkoudakis2022}. In addition, as PAH molecules are predominantly pumped by FUV photons copiously emitted by massive new-born stars, their emission can effectively be used to trace the star-formation rate (SFR) in galaxy-wide and spatially resolve scales in the case of metal-rich and dust-rich galactic environments \citep[e.g.][]{Calzetti2007, Shipley2016, Maragkoudakis2018b}. Similarly to their association with warm dust emission (i.e. SFR), PAHs are shown to also correlate with the diffuse cold dust, offering an alternative proxy of the molecular gas in star-forming galaxies \citep[e.g.] []{Regan2006, Cortzen2019}.

While PAH emission features are ubiquitous and present in almost all objects, their detailed spectral characteristics such as their relative strengths, central wavelengths, and spectral shapes are known to vary among (different type of) sources or even within a given source \citep[e.g.][]{Joblin1996, Sloan1997, Hony2001, Peeters2002, vanDiedenhoven2004, Berne2007, Smith07b, Sloan2007, Galliano2008b, Boersma2013, Sloan2014, Shannon2016, Stock2016, Peeters2017}. Such variations depict the diversity among the chemical structure, charge state, size distribution, and number density of the PAH populations within different environments, and are directly linked to and shaped by the local physical conditions of their surroundings. A comprehensive characterization of these variations is crucial for the robust determination of the local physical conditions of PAH environments, as well as the molecular identification of the emitters.

The formation mechanisms and subsequent processing that PAHs undergo within their environments will ultimately define their chemical and physical characteristics, and consequently shape the mid-IR spectral properties of the sources in which they reside. Condensation of metals in the ejecta of carbon-rich asymptotic giant branch (AGB) stars and core-collapse type II supernovae from massive stars are considered the prime sources for carbon dust production in galaxies and likely a supplier of interstellar PAHs \citep[e.g.][]{Latter1991,Todini2001}. \citet*{Jones1996} suggested that PAHs can also form via shattering in grain--grain collisions, while photoprocessing of very small grains (VSGs) by UV radiation is an additional proposed route of PAH formation \citep[e.g.][]{Cesarsky2000, Pilleri2012}. In addition, photoprocessing under UV radiation of the aliphatic bonds in HAC grains, can lead to the formation of aromatic bonds, and therefore to the formation of PAH molecules \citep{Jones2009, Sloan2007, Duley2015}. 

A combination of PAH processing in the form of fragmentation or destruction by (supernova) shocks \citep*{OHalloran2006} as well as by harder and/or more intense radiation fields \citep{Hony2001, Gordon2008, Berne2012, Croiset2016, Peeters2017, Knight2021}, or even a time lag between the formation of PAHs in AGB stellar atmospheres and their injection into the ISM during their post-AGB phase \citep*{Galliano2008a}, can have a direct impact on the number densities and the size-distribution of PAHs, especially in low-metallicity environments. \citet*{Micelotta2010} demonstrated that interstellar PAHs with a number of carbon atoms \Nc{} $ = 50 $ do not survive in shocks with velocities greater than 100 km s$ ^{-1} $, while larger PAHs (\Nc{} $ = 200 $) can be destroyed in shocks with velocities greater than 125 km s$ ^{-1} $. On the other hand, in shocks with velocities lower than 100 km s$ ^{-1} $, PAHs are not completely destroyed although they undergo a 20--40\% fraction loss of the carbon atoms. Furthermore, \cite{Hunt2010}, who studied a sample of blue compact dwarf (BCD) galaxies, presented evidence of larger relative intensities of the 8.6 and 11.2 \micron{} features, with respect to the other prominent PAH bands, which was attributed to the destruction of smaller PAHs in the harsher and more intense radiation fields of the low-metallicity BCD galaxies.  Evidence of the photochemical evolution of PAHs under UV processing has been presented through the examination of the variations in the 11.2/3.3 band ratio--known to track PAH size \citep{Schutte1993, Ricca2012}--with distance from the star \citep{Croiset2016, Knight2021}.  Alternatively, \cite{Sandstrom2010, Sandstrom2012} proposed that PAH molecules in the low-metallicity environment of the Small Magellanic Cloud (SMC) can either be forming in smaller sizes inside dense molecular cloud regions, or large PAHs may not be efficiently forming, and the small sized PAHs formed are more susceptible to destruction, explaining the PAH deficit observed in the SMC. However, the PAH size determination in the previous work was based on the 6.2/7.7 PAH band ratio which has been demonstrated to be an insensitive PAH size tracer \citep{Maragkoudakis2020}. 

In this paper, we utilized the NASA Ames PAH IR Spectroscopic Database\footnote{\href{https://www.astrochemistry.org/pahdb/}{https://www.astrochemistry.org/pahdb/}.} (PAHdb; \citealt{Bauschlicher2010};  \citealt{Boersma2014a}; \citealt{Bauschlicher2018}) to understand and quantify the variations in the spectra of PAH molecules under different size distributions, reflecting various processing/formation scenarios. This paper is structured as follows: Section \ref{sec:Sample} presents the selection of PAH molecules used in our analysis. A description of the emission model, calculations of spectra, and spectral decomposition methods is given in Section \ref{sec:Spectra}. We present our results on the PAH spectral variations for different PAH size distributions in Section \ref{sec:Results} and discuss their dependency on the spectral decomposition method used. A summary of our results and conclusions is given in Section \ref{sec:Summary}.

	\begin{figure*}
		\begin{center}
			\includegraphics[keepaspectratio=true,scale=0.5]{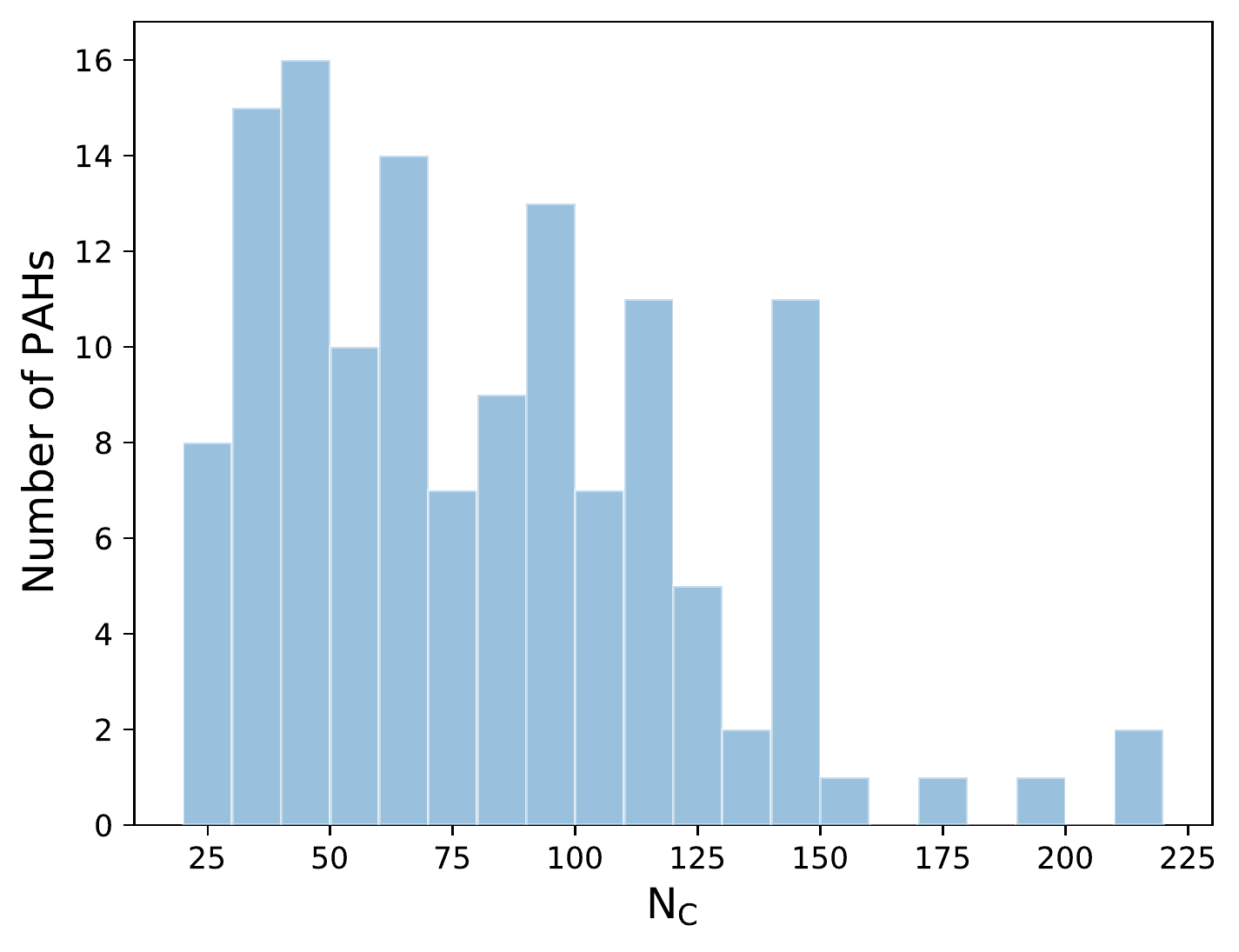}
			\includegraphics[keepaspectratio=true,scale=0.5]{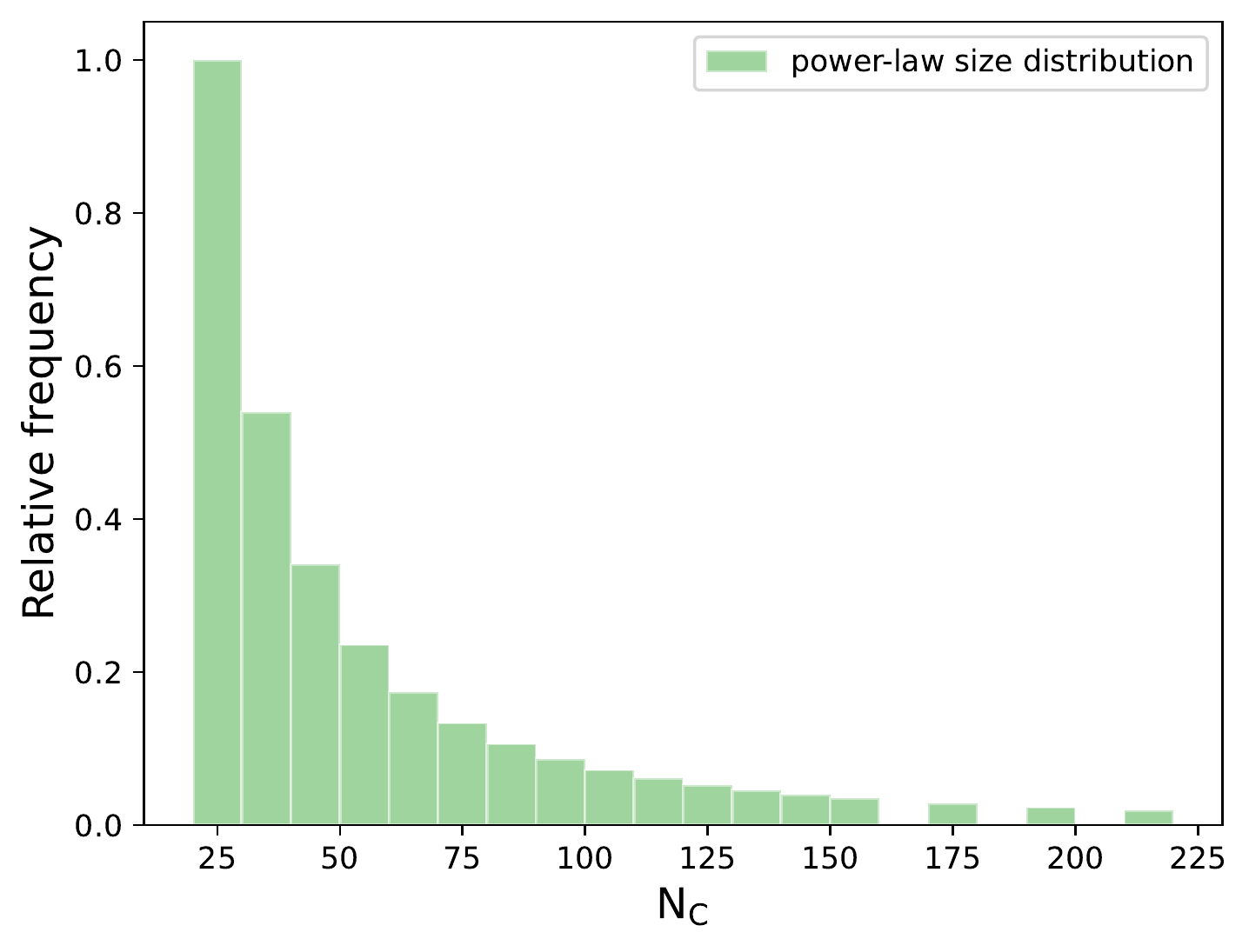}
			\caption{\textbf{Left:} Histogram of the size distribution of PAHs in the sample. \textbf{Right:} Distribution of the sample's \Nc{} (shown in the left-panel), when assuming the power-law distribution of \protect\cite{Schutte1993}, with the frequency normalized relative to the frequency in the first bin.}
			\label{fig:PAHdistributions}
		\end{center}
	\end{figure*}
	
\section{Sample} \label{sec:Sample}

In this work we used the sample of PAH molecules presented in \citet*[][hereafter \citetalias{Maragkoudakis2020}]{Maragkoudakis2020}. The selection criteria are briefly summarized as follows. Utilizing version 3.20 of the NASA Ames PAHdb (v3.20 was extended to include neutral and cationic pairs of the straight-edge molecules presented in \citealt{Ricca2018}), pairs of neutral and singly charged cationic molecule pairs were selected that: (i) contain more than 20 carbon atoms assumed to survive typical radiation field conditions of interstellar environments \citep[e.g.][]{Allain1996}; (ii) have no nitrogen, oxygen, magnesium, or iron atoms, in order to examine a consistent sample of ``pure" PAH populations; (iii) have solo C-H bonds to ensure the inclusion of molecules producing the key 11.2 and 11.0  \micron{} emission features due to C--H out-of-plane modes of solo hydrogens in neutral and cationic species respectively. Given the previous criteria, the final sample consists of 266 molecules (133 neutral--cationic pairs) with a size distribution between 22 and 216 carbons (Figure \ref{fig:PAHdistributions}, left panel). The PAH sample formulas and their PAHdb unique identifiers (UIDs) are given in Table \ref{tab:UIDsv320}. We emphasize that the analysis and diagnostic diagrams presented in this paper are for a sample of pure PAHs only (i.e. the sample does not include e.g. hetero-atoms).

\section{Model and spectra} \label{sec:Spectra}

We followed a similar approach to \citetalias{Maragkoudakis2020} for the emission model application and generation of the emission spectrum of the individual PAH molecules in the sample. Specifically, we considered PAHs subjected to the interstellar radiation field (ISRF) of \citet*{Mathis1983}, where the entire radiation field spectrum was accounted for instead of the average absorbed photon energy for each PAH (see Appendix A in \citetalias{Maragkoudakis2020} for a detailed description). The entire emission cascade was taken into account for the calculation of the PAH emission spectrum, while no subsequent redshift was applied regarding anharmonicity corrections in accordance to \cite{Mackie2018}. Finally, the PAH spectral emission bands were convolved using Lorentzian line profiles with a full width at half-maximum of 15 cm$^{-1}$. Following \citetalias{Maragkoudakis2020}, aside from the purely neutral and cationic molecule spectra, we synthesized spectra that correspond to populations with a mixture of neutral and cationic PAH components. The synthesized spectra were created by combining the individual spectra of neutral--cation molecule pairs with relative neutral--cation contributions of: (i) 75--25 per cent (N75C25), (ii) 50--50 per cent (N50C50), and (iii) 25--75 per cent (N25C75).

\begin{figure*}
		\begin{center}
			\hspace*{-0.5cm}\includegraphics*[keepaspectratio=true,scale=0.48]{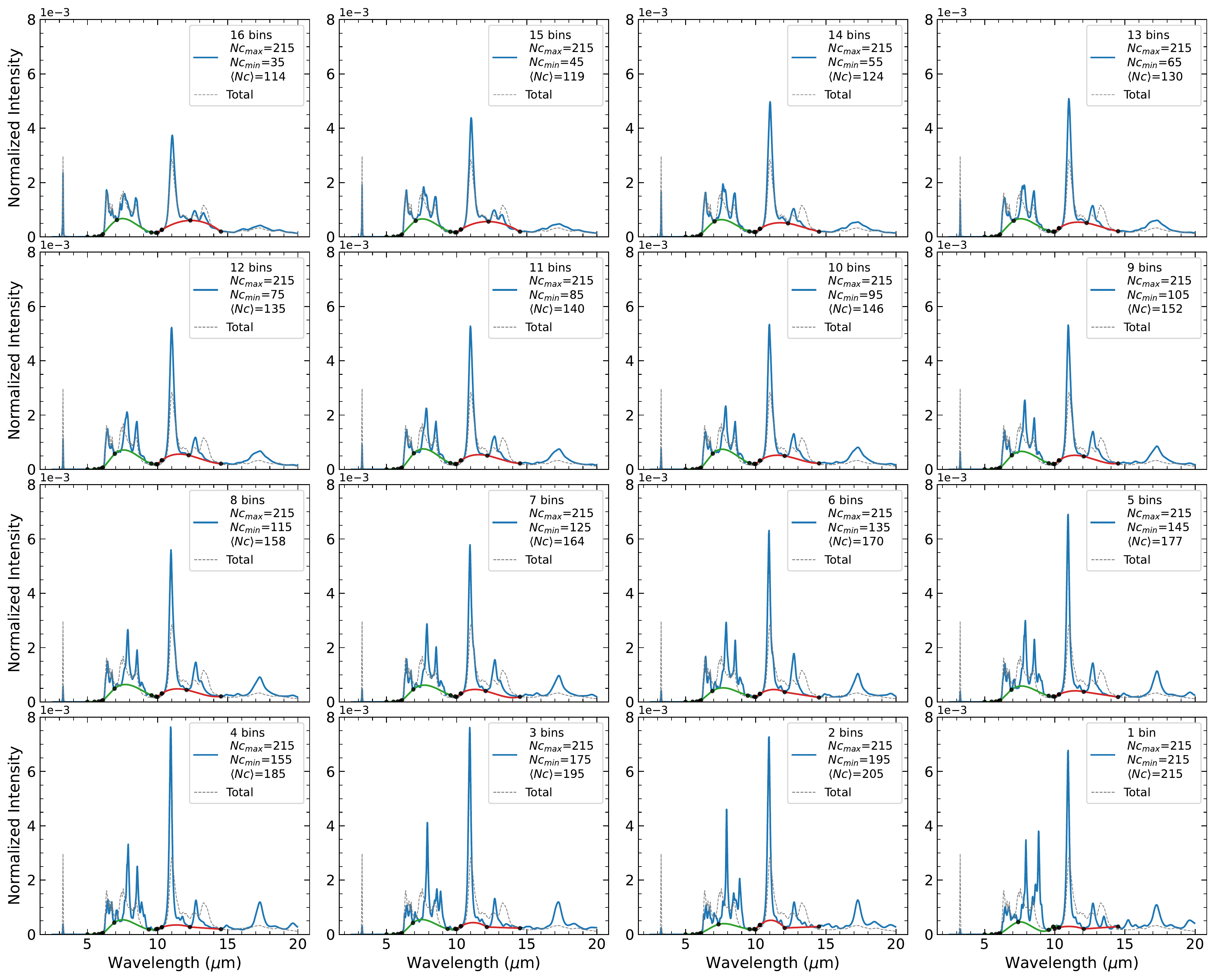}
			\caption{Successive subtraction of individual averaged PAH spectra of a given bin from the total synthesized spectrum (shown with a dashed gray line in all panels), starting from the smallest PAHs bin (SPR) for the N50C50 case. Each spectrum (including the total spectrum) is given in erg/cm$^{-1}$, has spectral resolution of $\Delta_{\nu}=2$ cm$^{-1}$, and is normalized to its total flux. The legend describes the number of bins used to synthesize the spectrum in each panel (shown in blue), the smallest number of carbon atoms in the spectrum (N$_{\mathrm{Cmin}}$), and the abundance-weighted average \Nc{} (\AvgNc). The green and red lines are the local continua in the 5-10 \micron{} and 10--15 \micron{} range respectively, defined with the spline method to selected anchor points (black points).}
			\label{fig:SPR_binspec}
		\end{center}
	\end{figure*}

\begin{figure*}
		\begin{center}
			\hspace*{-0.5cm}\includegraphics*[keepaspectratio=true,scale=0.48]{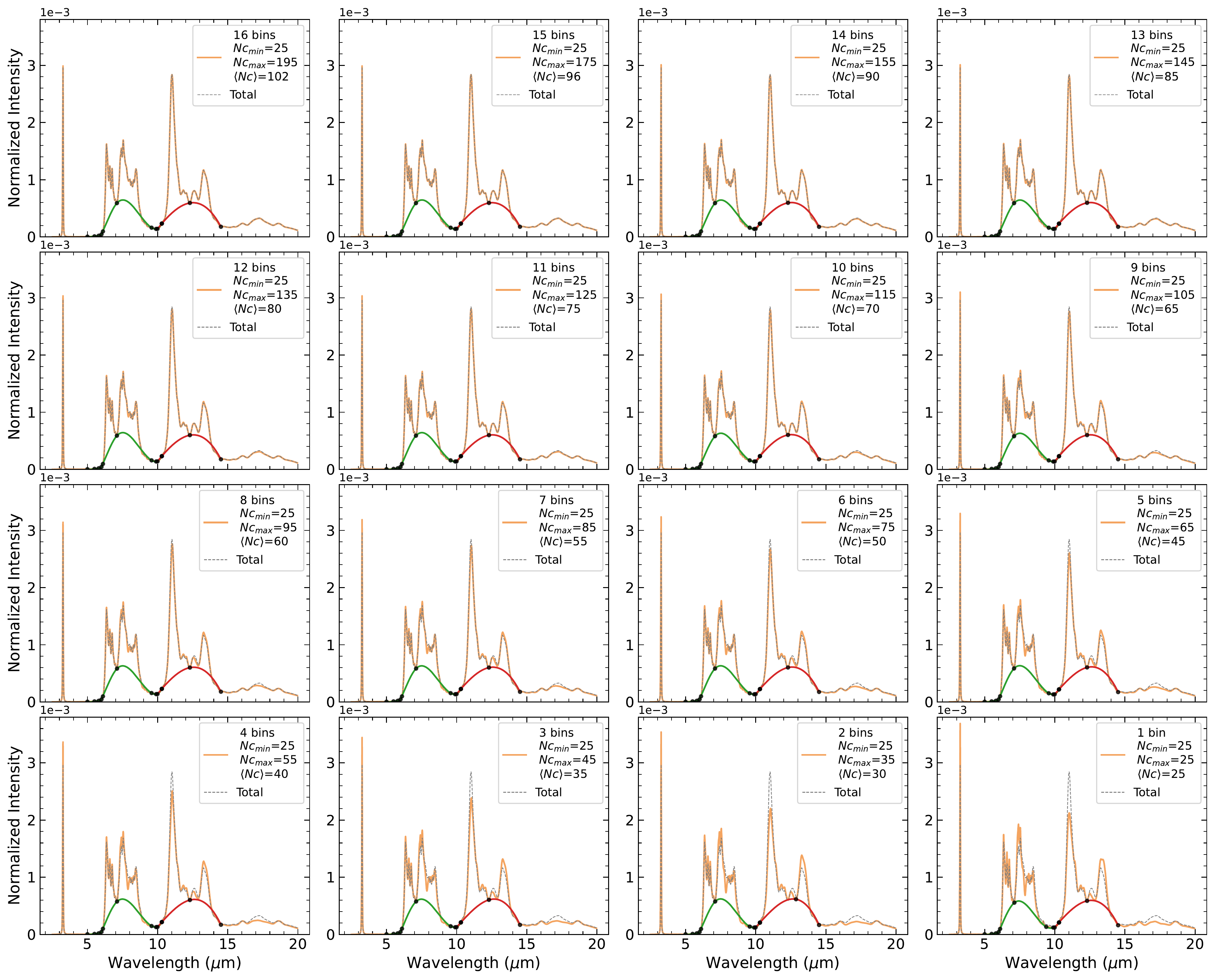}
			\caption{Successive subtraction of individual averaged PAH spectra of a given bin from the total synthesized spectrum (shown with a dashed gray line in all panels), starting from the largest PAHs bin (LPR) for the N50C50 case. Each spectrum (including the total spectrum) is given in erg/cm$^{-1}$, has spectral resolution of $\Delta_{\nu}=2$ cm$^{-1}$, and is normalized to its total flux. The legend describes the number of bins used to synthesize the spectrum in each panel (shown in orange), the largest number of carbon atoms in the spectrum (N$_{\mathrm{Cmax}}$), and the abundance-weighted average \Nc{} (\AvgNc).The green and red lines are the local continua in the 5-10 \micron{} and 10--15 \micron{} range respectively, defined with the spline method to selected anchor points (black points).}
			\label{fig:LPR_binspec}
		\end{center}
	\end{figure*}

In this work, we examine two scenarios of PAH processing or equivalently PAH formation: \textbf{(1)} small PAHs are being destroyed (or equivalently large
PAHs are more efficiently formed), to which we refer to as SPR (small PAHs removed), resulting in a larger average size of the PAH population; \textbf{(2)} large PAHs are being destroyed (or equivalently small PAHs are more efficiently formed), to which we refer to as LPR (large PAHs removed) resulting in a decrease of the average PAH size of the PAH population. The replication of the different scenarios in successive steps, followed by the examination of their spectral characteristics is described as follows. We divide the PAH spectra in 20 bins based on their \Nc{} distribution (\Nc$_{\mathrm{min}}=22$, \Nc$_{\mathrm{max}}=216$), with the bin size set to $\mathrm{\Delta N_{C}}=10$ and the central bin values being the average of the bin edges, i.e. the smallest bin corresponding to a central value of 25 \Nc{} and the largest bin to a central value of 215 \Nc. In each bin, we calculated the average spectrum from the combination of the individual spectra within that bin, and subsequently weighting the resulting averaged spectrum assuming the power-law size distribution of \cite{Schutte1993}:
\begin{equation}
    \frac{dN_{PAH}}{N_{H}}=B_{C}N_{C}^{-\beta-1}dN_{C},
\end{equation}
 where $dN_{PAH}/N_{H}$ is the total number of PAHs per interstellar H atom containing between $N_{C}$ and $N_{C}+dN_{C}$ carbon atoms, $B_{C}=1.24 \times 10^{-6}$, $\beta=0.833$, and assumed hydrogen column density $N_{H}=1.9 \times 10^{21}$ cm$^{-2}$ (Figure \ref{fig:PAHdistributions}, right panel). Finally, the total spectrum is calculated from the combination of the weighted averaged spectrum of each bin. To reproduce the different SPR and LPR scenarios described previously, we successively subtracted from the total spectrum the individual averaged and weighted spectrum of each bin, starting from the smallest \Nc{} bin for the SPR case and gradually subtracting the adjacent bins as we move to larger \Nc{} bins, or similarly for the LPR case by starting from the largest bin and gradually subtracting the adjacent bins as we move to smaller \Nc{} bins. The resulting spectra in each step and scenario for the N50C50 case are presented in Figures \ref{fig:SPR_binspec}, and \ref{fig:LPR_binspec}, while the spectra for the remaining charge balances are provided as online material. 

\subsection{Spectral decomposition and emission line measurements} \label{sec:EmissionLines}

The intensities of the prominent PAH emission bands at 3.3, 6.2, 7.7, 8.6, and 11.2\footnote{The emission feature due to the solo CH out-of-plane bending mode is referred to as the 11.2 \micron{} feature for the case of neutral molecules, the 11.0 \micron{} feature for cations, and the (11.2+11.0) feature for the N75C25, N50C50, and N25C75 cases (see \citetalias{Maragkoudakis2020} for a detailed description).}, were measured using two methods. The first method measures the band fluxes as the sum of fluxes between predefined wavebands following \citetalias{Maragkoudakis2020}, without separating the contribution of plateau emission. In the second method, we fitted and subtracted the plateau emission underneath the PAH features (see Figures \ref{fig:SPR_plr} and \ref{fig:LPR_plr}) before measuring their intensities. Specifically, we defined two local continua: \textit{(1)} under the 6.2 \micron, 7.7 \micron{} complex, and 8.6 \micron{} emission features, in the 5--10 \micron{} wavelength range; \textit{(2)} under the 11.2 and 12.7 \micron{} features, in the 10--15 \micron{} wavelength range. In the 5--10 \micron{} region, the plateau was defined by fitting a spline function to selected anchor points. In the 10--15 \micron{} region, a spline or a combination of a spline and a line was used to define the plateau. The individual PAH emission features on top of the 5--10 \micron{} and 10--15 \micron{} plateaus were estimated by subtracting the plateau emission from the integrated flux of each PAH feature as described and calculated in the first method. 

\section{Results and Discussion} \label{sec:Results}

We describe the variance and contribution of the individual PAH features to the total PAH spectrum for the different cases of PAH processing or formation in Section \ref{sec:pahfeat}. The plateau emission characteristics in the SPR and LPR cases are examined in Section \ref{sec:plateau}, and the relative PAH intensity variations are discussed in Section \ref{sec:RelIntens}. Finally, we investigate the behavior of the different ionization fractions in Section \ref{sec:IonFrac} and present new PAH charge -- size diagnostic diagrams in Section \ref{sec:new_grid}. 

	\begin{figure*}
		\begin{center}
			\hspace*{-0.1cm}\includegraphics[keepaspectratio=true,scale=0.6]{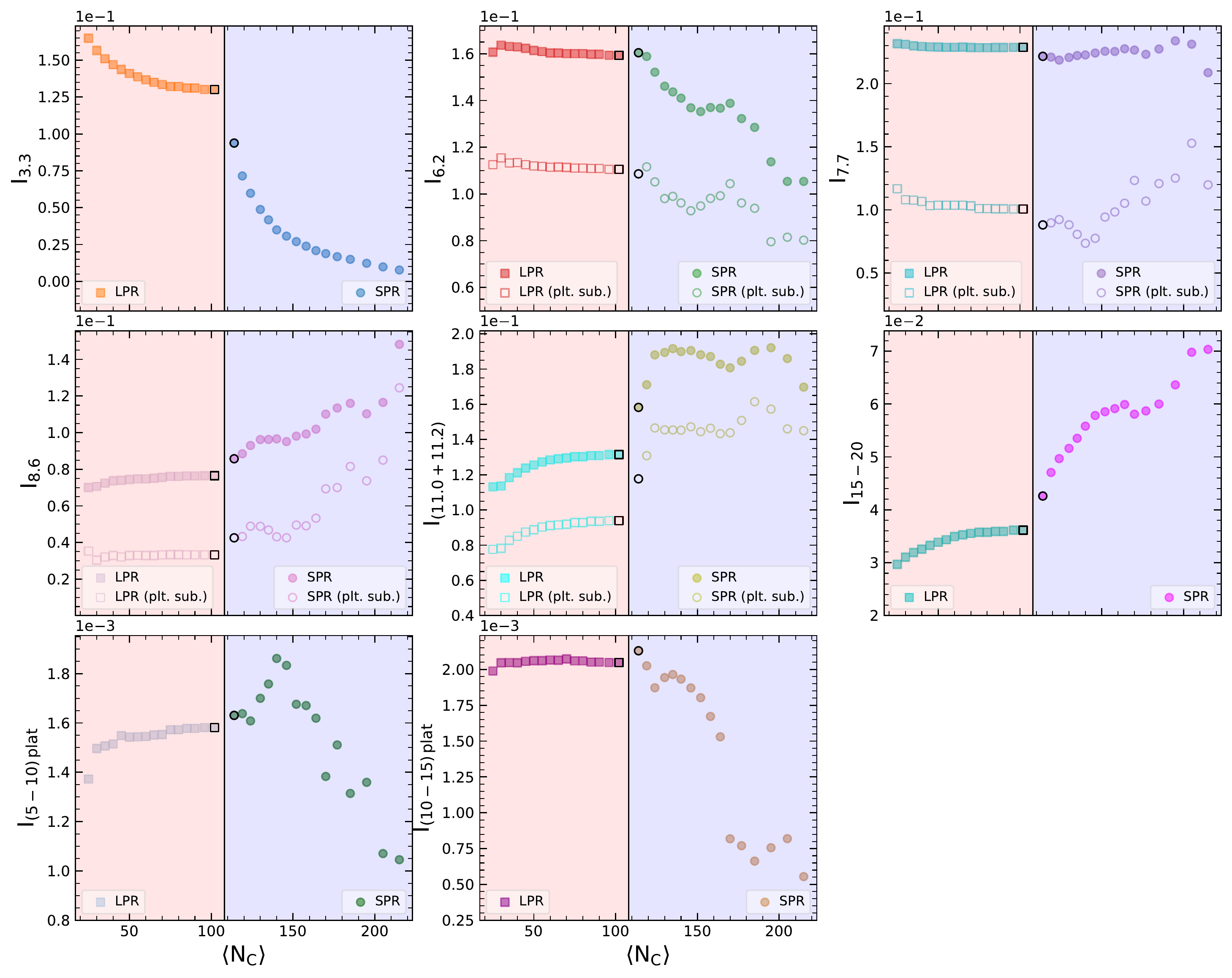}
			\caption{Intensity variations of PAH features and plateaus normalized to the total flux of each spectrum for the N50C50 case, as a function of \AvgNc, for the small PAHs removed (SPR)  and large PAHs removed (LPR) cases. Open circles and squares, where present, indicate plateau subtracted fluxes for the SPR and LPR cases respectively. Black edged circles and squares indicate the initial resulting spectrum after the removal of the first bin's spectrum in each case (SPR and LPR) from the total spectrum. All intensities are normalized to the total flux of each individual spectrum.}
			\label{fig:Int_AvgNc_N50C50}
		\end{center}
	\end{figure*}

\subsection{Variations of the PAH features' contribution to the total PAH emission.} \label{sec:pahfeat}

Figure \ref{fig:Int_AvgNc_N50C50} presents the intensity variations of the main PAH features (with and without plateau subtraction) normalized to the total flux of the spectrum, as a function of the abundance-weighted average number of carbon atoms (\AvgNc{}) for each step of the N50C50 SPR and LPR cases. We discuss first the SPR/LPR cases without plateau subtraction. In the SPR case, the contributions of the 3.3 and 6.2 \micron{} features to the total spectrum are progressively decreasing as smaller PAHs are gradually removed from the total spectrum. This is in accordance with the fact that smaller PAHs are contributing most of their emission to shorter wavelengths \citep*{ATB1989, Schutte1993}. On the other hand, the contribution to the total flux of the 7.7 \micron{} complex, 8.6 \micron{} feature, as well as the sum of fluxes in the 15--20 \micron{} region (I$_{15-20}$), are all incrementally increasing as \AvgNc{} increases--with the 7.7 \micron{} complex showing a more subtle increase compared to the rest, which is likely justified considering that larger PAHs are predominately contributing to the emission at longer wavelengths. However, the 11.2 \micron{} emission feature exhibit a non-monotonic behaviour as smaller PAHs are removed from the total spectrum and consequently the \AvgNc{} increases. The 11.2 \micron{} feature contribution to the total spectrum increases after the subtraction of the first two bins, while it remains relatively constant until \AvgNc{} $= 195$, with only small fluctuations, and then decreases for \AvgNc{} $> 195$. 
Among the individual PAH bands, the 3.3 \micron{} band has the highest dependence on \AvgNc, in accordance with \citetalias{Maragkoudakis2020}, having the largest dynamic range in intensity as a function of \AvgNc. 


The profiles for the majority of PAH emission bands (Figure \ref{fig:SPR_binspec}) present minimal to moderate variations, with the exception of the 7.7 \micron{} complex and the 8.6 \micron{} feature which show more profound variations. Specifically, the 7.7 \micron{} profile is progressively redshifted with increasing \AvgNc{} (see Appendix \ref{app:77_region}, Figure \ref{fig:SPR_77}), also noted by \cite{Bauschlicher2008} and \cite{Shannon2019} for the case of individual PAHs of different sizes, while the 8.6 \micron{} feature is splitting into two emission components at 8.6 \micron{} and 8.9 \micron{} from \AvgNc{} $>135$ with comparable peak strengths at \AvgNc{} $=195$ and the ``red" emission component having higher peak flux in the last two \Nc{} bins. 


In the N50C50 LPR case, the intensity variations of the individual PAH features with respect to the total spectrum flux are less prominent, span a smaller range, and show a monotonic behavior compared to the SPR case (Figure \ref{fig:Int_AvgNc_N50C50}). The 3.3 and 6.2 \micron{} features increase with decreasing \AvgNc, while the 7.7 \micron{} complex and 8.6 \micron{} emission feature remain relatively constant with the 7.7 \micron{} feature marginally increasing with decreasing \AvgNc{}, and the 8.6 \micron{} feature slightly decreasing. We note that the 7.7 \micron{} feature profile shows minor variation with \AvgNc{} (see Appendix \ref{app:77_region}, Figure \ref{fig:LPR_77}). The 11.2 \micron{} feature intensity progressively decreases towards lower \AvgNc{} along with the I$_{15-20}$. Similar to the SPR case, the 3.3 \micron{} band has the highest dependence with \AvgNc{}, in accordance with \citetalias{Maragkoudakis2020}, with the largest dynamic range in intensity as a function of \AvgNc{} among the individual PAH bands. The PAH emission band profiles do not show variation in shape or shift, as opposed to the SPR case. 

When a plateau component is accounted for and subtracted from the corresponding integrated emission of a PAH bandpass (see Section \ref{sec:plateau}), the resulting PAH emission feature will naturally have a lower intensity compared to the case where the entire bandpass intensity is attributed to the PAH feature. The intensity variations of the 6.2, 7.7, 8.6, and 11.2 \micron{} feature contribution to the total emission when a plateau emission is subtracted are also displayed in Figure \ref{fig:Int_AvgNc_N50C50}. In both the SPR and LPR cases, all features but the 7.7 \micron{} complex in the SPR case present similar patterns to the intensity variations when the entire bandpass emission is measured and attributed to the corresponding PAH feature. For the 5--10 \micron{} plateau, this is indicative that the fraction of plateau emission underneath the 6.2 and 8.6 \micron{} feature is less significant compared to the plateau emission underneath the 7.7 \micron{} complex.  

\subsection{Plateau emission} \label{sec:plateau}

The plateau emission at 5--10 \micron{} and 10--15 \micron{} was modeled using a spline function on a set of anchor points (or a combination of a spline and a line; Section \ref{sec:EmissionLines}), and their respective intensities were estimated by calculating the spline integrals. Figure \ref{fig:Int_AvgNc_N50C50} also presents the plateau intensity variations as a function of the total spectrum flux for the N50C50 SPR and LPR cases. The 5--10 \micron{} plateau intensity shows a bi-modal behavior in the SPR case, increasing up to \AvgNc{} $=140$ and then declining for larger \AvgNc{} values. In the LPR case the 5--10 \micron{} plateau intensity slightly decreases as the \AvgNc{} is reduced. The 10--15 \micron{} plateau intensity has an overall declining trend with small fluctuations present in the SPR case, while remains almost constant in the LPR case. For \AvgNc{} $> 164$, the 10--15 \micron{} plateau in the SPR case was fitted with a combination of a spline (10--12 \micron) and a line (12--15 \micron{}), due to the flattening of the plateau emission in the 12--15 \micron{} region. This translates to a \textit{jump} in the calculated 10--15 \micron{} plateau intensity (Fig. \ref{fig:Int_AvgNc_N50C50}, bottom right panel) when going from \AvgNc{} $= 164$ (spline only) to \AvgNc{} $= 170$ (spline and line), however the plateau fluctuations for \AvgNc{} $> 170$ are successfully captured as they are calculated and described in a self-consistent manner. Overall, the relative variations in the plateaus emission among the SPR and LPR cases are suggestive that plateaus are dominated by smaller / medium-sized PAHs. The plateaus in these cases are produced from small variations in the peak positions of the emission features of the individual molecules, which will add up when synthesizing the spectrum within a bin and form the observed plateau. The variation in the features' peak positions is seen mostly in small to medium-sized PAHs in our sample, explaining the higher variation of the plateau emission in the SPR case. Since the sample in the present study does not include PAH clusters, it is not clear whether the observed plateau emission in 5--10 \micron{} and 10--15 \micron{} \citep[e.g.,][]{Tielens2008} underlying the AIB features are fully explained by the presence and/or absence of the small-intermediate PAHs. However, the present results indicate that not just the PAH clusters but also the small-intermediate PAHs are responsible, at least, for a part of the plateau emission observed in 5--10 \micron{} and 10--15 \micron.

\subsection{PAH relative intensities variations} \label{sec:RelIntens}

\begin{figure*}
		\begin{center}
			\includegraphics[keepaspectratio=true,scale=0.6]{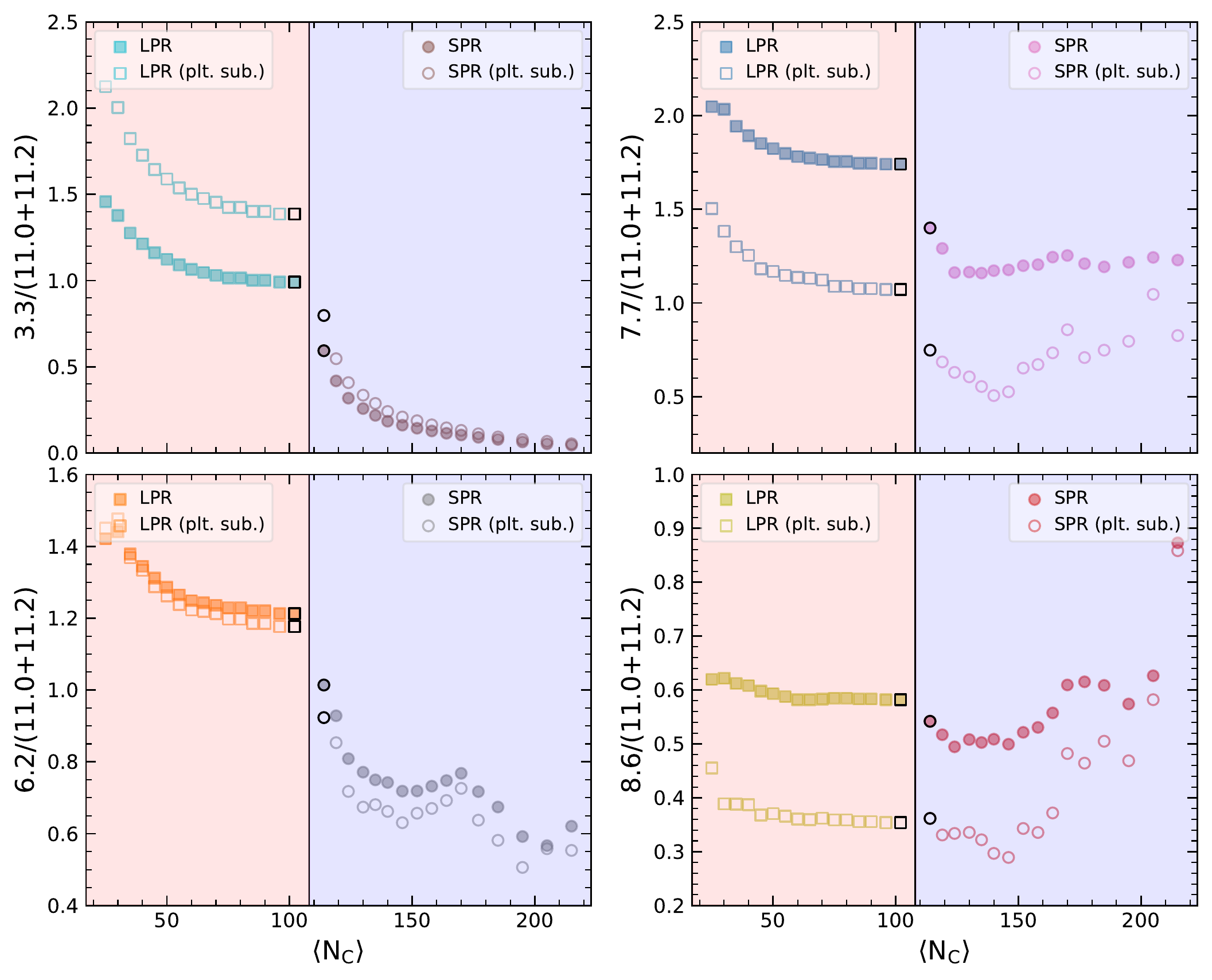}
			\caption{PAH intensity variations relative to the (11.2+11.0) \micron{} emission as a function of \AvgNc{} for the N50C50 SPR and LPR cases. Open circles and squares indicate plateau subtracted intensity ratios for the SPR and LPR cases respectively. Black edged circles and squares indicate the intensity ratios of the initial resulting spectrum after the removal of the first bin's spectrum in each case (SPR and LPR) from the total spectrum.}
			\label{fig:intratio_AvgNc_N50C50}
		\end{center}
	\end{figure*}

We examined the behaviour of relative PAH intensity ratios in the SPR and LPR cases with and without a plateau emission subtraction (Figure \ref{fig:intratio_AvgNc_N50C50}). Specifically, we investigated the intensity variations in the 3.3/(11.2+11.0), 6.2/(11.2+11.0), 7.7/(11.2+11.0), and 8.6/(11.2+11.0) band ratios. 

The intensity of the 3.3/(11.0+11.2) ratio gradually decreases in the SPR case and increases in the LPR case. Because the (11.0+11.2) flux remains relatively constant in the SPR case (although increases in the first bins; see Section \ref{sec:pahfeat}) and gradually decreases in the LPR case, the behaviour of the 3.3/(11.0+11.2) intensity ratio is governed by the larger 3.3 \micron{} variations. This is in accordance with \citetalias{Maragkoudakis2020}, who demonstrated that the 3.3 \micron{} band has the strongest dependence with \Nc, and is the main driver behind correlations of the intensity ratio of the solo CH out-of-plane to CH stretching mode (i.e. (11.0+11.2)/3.3) with \Nc, for neutral, cationic, and anionic PAHs. Furthermore, while the relative difference between the standard (no plateau subtracted) and plateau subtracted 11.0+11.2 \micron{} emission in the SPR and LPR case is comparable, the lower values and steep decrease of the 3.3 \micron{} intensity in the SPR case smooths out the difference between the plateau vs no plateau subtracted 3.3/(11.2+11.0) ratios, as opposed to the LPR case. In both SPR and LPR cases the 3.3/(11.2+11.0) intensity ratios are higher when a plateau emission is subtracted compared to non-plateau-subtracted ratios, as a result of the lower 11.0 and 11.2 \micron{} fluxes in the plateau subtracted cases. 


The 6.2/(11.2+11.0) intensity ratio, similarly to the 3.3/(11.2+11.0) ratio, decreases with increasing \AvgNc{} in the SPR case, and increases with decreasing \AvgNc{} in the LPR case. In the SPR case, the drop in the intensity is driven by the 6.2 \micron{} feature flux, with the variation pattern controlled by the 11.2+11.0 variations due to the monotonic decrease of the 6.2 \micron{} emission. In the LPR case, the intensity increase and variation shape of the 6.2/(11.2+11.0) ratio is regulated by the 11.2+11.0 behavior. When examining plateau-subtracted variations, both the 6.2 and 11.2+11.0 \micron{} features are affected by the corresponding fraction of their respective plateau emission, with the 10.0--15.0 \micron{} plateau having the largest variation in intensity. In both SPR and LPR cases, the 6.2/(11.2+11.0) intensity variations with and without plateau subtraction have comparable fluxes and similar trends. In this case, the non-plateau-subtracted ratios have marginally higher intensities than the plateau-subtracted ones, with as only exception the spectra in the last two bins of the LPR case.

The LPR intensities of the 7.7/(11.2+11.0) and 8.6/(11.2+11.0) ratios increase with decreasing \AvgNc{} and these are mainly driven by the 11.2+11.0 \micron{} variations which progressively decrease with decreasing \AvgNc, as the 7.7 and 8.6 \micron{} contributions to the total spectrum flux remain fairly insensitive to \Nc{}. Similar variation patterns are seen in the plateau-subtracted case at lower intensities. In the SPR case, the 7.7/(11.2+11.0) ratio remains fairly constant as the respective 7.7 and (11.2+11.0) \micron{} feature intensities have similar variations, especially for \AvgNc{} $>150$, and cancel out. The plateau-subtracted 7.7/(11.2+11.0) variations have similar fluctuations to the plateau-subtracted 7.7 \micron{} flux which are of higher magnitude than the corresponding (11.2+11.0) \micron{} variations. The 8.6/(11.2+11.0) ratio intensity in the SPR case has a generally increasing tendency with increasing \AvgNc{}, with fluctuations being the product of the respective variance in the 8.6 and 11.2+11.0 \micron{} fluxes. Similarly, the plateau-subtracted 8.6/(11.2+11.0) intensity is increasing with increasing \AvgNc{}, with the variance driven by the plateau-subtracted 8.6 \micron{} flux. For both 7.7/(11.2+11.0) and 8.6/(11.2+11.0) ratios, the plateau-subtracted fluxes are lower compared to the respective non-plateau-subtracted fluxes, and the relative flux difference is higher compared to the 3.3/(11.2+11.0) and 6.2/(11.2+11.0) ratios.

\subsection{Spectra of different ionization fractions}  \label{sec:IonFrac}

 The PAH intensities and plateau variations presented for the N50C50 SPR and LPR cases (Section \ref{sec:pahfeat}; Figure \ref{fig:Int_AvgNc_N50C50}) also apply when examining the spectra of the different ionization fractions, i.e., the purely neutral, N75C25, N25C75, and purely cationic spectra (Appendix \ref{sec:Int_AvgNc_various}, Figures \ref{fig:Int_AvgNc_Neutral}-\ref{fig:Int_AvgNc_Cation}). Aside from differences in the absolute intensity of the respective bands or plateaus, we identify similar trends for the spectra of different ionization fractions in both their SPR and LPR cases as with the N50C50 case, with the only exception that of the 6.2 and 7.7 \micron{} bands in the purely neutral PAH spectra. Attributed predominately to ionized PAHs, the 6 to 9 \micron{} emission is intrinsically an order of magnitude larger in ionized PAHs compared to that of neutral PAHs \citep[e.g.][]{Bauschlicher:vlpahs1, Bauschlicher:vlpahs2, Ricca2012, Ricca2018, Ricca2019}. This is due to the fact that the intensity of the bands correlates with the dipole derivatives and ionized PAHs have larger dipole moments than neutral PAHs.
 
 The intensity of the 6.2 \micron{} PAH band for all cases with ionized PAHs (i.e., all but the purely neutral case) overall decreases with \AvgNc{} as the amount of charge per carbon atom decreases with size. For the Neutral case, PAH structure affects the intensity of the 6.2 \micron{} PAH band. Small PAHs have a significant number of structures that are mostly non-compact with eroded edges with less intense 6.2 \micron{} bands. Removing small PAHs results in an overall increase of the intensity of the 6.2 \micron{} PAH band. Large compact PAHs have similar intensities of the 6.2 \micron{} PAH band whereas large PAHs with eroded edges have less intense 6.2 \micron{} PAH bands. 
 The intensity of the 7.7 \micron{} PAH band is related to the charge distribution on the C-H bonds and shows a slight increase with \AvgNc{} for the Cation case and a slight decrease for the Neutral case.
 
\subsection{PAH charge -- size diagnostics for \AvgNc, $\mathrm{N_{C\,max}}$, and $\mathrm{N_{C\,min}}$}  \label{sec:new_grid}

\begin{figure*}
	    \begin{center}
			\hspace*{-0.5cm}\includegraphics[keepaspectratio=true,scale=0.38]{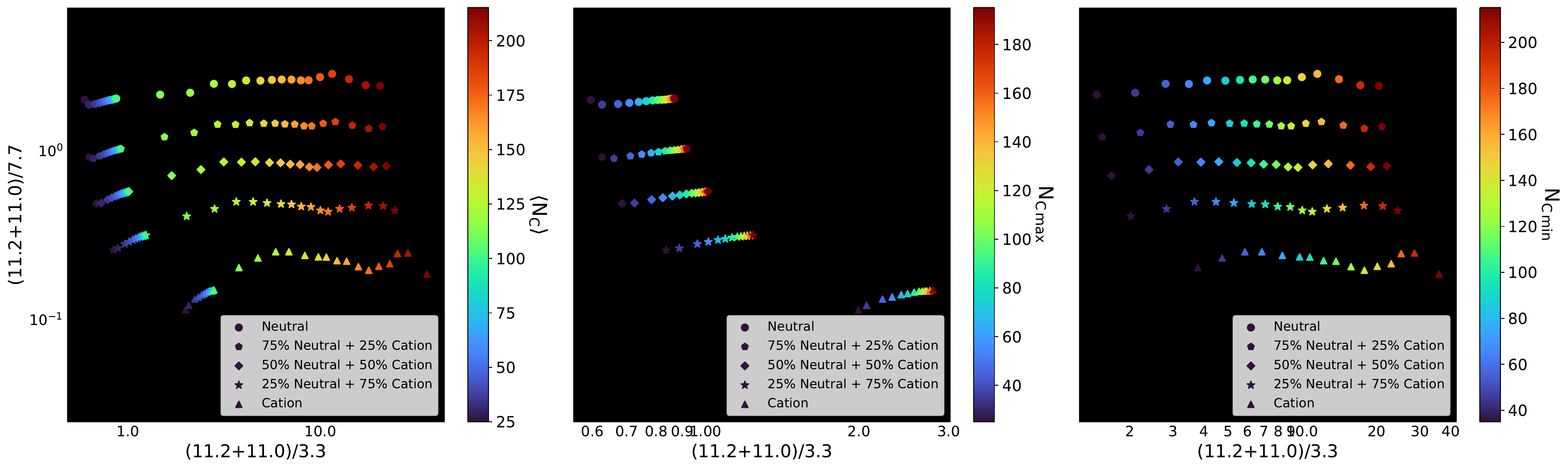}
			\caption{The PAH charge--size space generated from the spectra of each step of the LPR and SPR cases. Different shapes correspond to different ionization fractions as described in Section \ref{sec:Spectra}. \textbf{Left panel}: LPR and SPR spectra color-coded on \AvgNc; \textbf{Middle panel}: LPR spectra color-coded on $\mathrm{N_{C\,max}}$; \textbf{Right panel}: SPR spectra color-coded on $\mathrm{N_{C\,min}}$.}
			\label{fig:grid}
	    \end{center}
	\end{figure*}

Considering the spectra produced in the SPR and LPR analysis (Figures \ref{fig:SPR_binspec}-\ref{fig:LPR_binspec}) as PAH spectra of individual sources of different \Nc{} and ionization fractions, we are able to construct a PAH charge -- size space (Figure \ref{fig:grid}, left panel), similar to \citetalias{Maragkoudakis2020} (their Figure 8). The main difference between the current and \citetalias{Maragkoudakis2020} PAH charge -- size space is that the synthesized PAH spectra in the SPR and LPR analysis have been weighted assuming the power-law size distribution of \cite{Schutte1993} with a minimum PAH size of \Nc$_{\mathrm{min}}=22$ and a maximum PAH size of  \Nc$_{\mathrm{max}}=216$. 
Thus, the PAH charge -- size space introduced in this work is based upon a more representative size distribution of PAHs within astrophysical sources, and due to the larger number of bins (points) provides a more detailed description of the variance in the PAH charge and size. In addition, the current and \citetalias{Maragkoudakis2020} PAH charge -- size space are constructed in a slightly different manner, i.e., the \citetalias{Maragkoudakis2020} space points were defined from bins of the 11.2(11.0)/3.3 values, whereas in this work the SPR/LPR spectra are defined within \Nc{} bins, with different number of bins in each case. While the \Nc{} and the 11.2(11.0)/3.3 ratio are correlated, their correlation is not linear (\citetalias{Maragkoudakis2020}), resulting in small fluctuations of the points in the two diagrams. Nonetheless, the two diagrams are consistent and provide equivalent information on the average charge state and size of PAHs in a source. 

Our current SPR and LPR analysis provides access to the maximum or minimum \Nc{} of PAHs present within a source, displayed in the middle and right panels of Figure \ref{fig:grid} with the PAH charge -- size space color-coded on the $\mathrm{N_{C\,max}}$ and $\mathrm{N_{C\,min}}$ respectively. The combined diagrams offer a comprehensive description on the average, minimum, and maximum \Nc{} present within a source. In all cases the non-plateau subtracted intensities were considered, following the analysis of \citetalias{Maragkoudakis2020}.

\section{Astrophysical implications}

The presence (or absence) of small PAHs appears to have the most direct impact on the shape, individual band strengths, and plateau emission of a source's PAH emission spectrum. As such, the mechanisms and environments that process--or form small PAHs, will have a determining role on the resulting PAH spectrum of a region or a source. The shift and profile shape variations of the 7.7 \micron{} complex in the SPR case (Figure \ref{fig:SPR_77}) can provide additional information to the PAH profile classification scheme developed by \cite{Peeters2002} (and later complemented by \citealt{vanDiedenhoven2004} and \citealt{Matsuura2014}), where the processing or formation of PAHs may induce a transition between the different profile ``classes" \citep[][]{Bauschlicher:vlpahs2, Shannon2019}. 

The variation in the relative PAH intensities, frequently used as proxies of the PAH size \citep[i.e. 3.3/(11.0+11.2); e.g. ][]{Schutte1993, Ricca2012, Croiset2016, Maragkoudakis2020, Knight2021}, and the PAH ionization fraction \citep[i.e. 6.2/(11.2+11.0), 7.7/(11.2+11.0), and 8.6/(11.2+11.0); e.g. ][]{Joblin1996, Sloan1997, Allamandola1999}--and thus the ionization parameter \citep{Galliano2008b, Stock2016, Knight2022, Maragkoudakis2022}, are mainly governed by the PAH feature with the highest dependence on \Nc. Among the individual PAH bands, the 3.3 \micron{} band shows the highest dependence on \Nc{}, in both the SPR and LPR cases, as well as a monotonic behavior, which translates to a well defined scaling between the 3.3/(11.0+11.2) ratio and \Nc{} in both scenarios, thus setting the 3.3/(11.0+11.2) ratio as an effective proxy for tracking PAH size (variations). 
Regarding the PAH ionization proxies, while all of them show variance with \AvgNc, the 7.7/(11.0+11.2) ratio presents the least amount of variance with the lowest dynamic range, considering the SPR and LPR cases as a whole, while being almost constant in the SPR case, rendering the 7.7/(11.0+11.2) ratio as the better choice for tracing PAH charge. 

In addition to being an efficient proxy for PAH size, the 3.3/(11.0+11.2) ratio's monotonic behavior and fully separated values between the SPR and LPR cases allows the characterization of the dominant scenario of processing or formation within a region or a source, given that the charge balance is known (i.e. knowing which case of the pure neutral, N75C25, N50C50, N25C75, or pure cationic PAH case to employ). This can help understand the connection between different local physical conditions and environments within a source (e.g., between the H\,\textsc{ii} regions in a galaxy) in an evolutionary manner.

\section{Summary and conclusions} \label{sec:Summary}

In this work, we examined the variations in the intensity of the main PAH bands, PAH band ratios, and plateau emission as a function of PAH size (i.e. the number of carbon atoms \Nc{}), utilizing version 3.20 of the NASA Ames PAHdb, considering a size distribution of PAHs and assuming two scenarios: From an initial synthesized \textit{total} PAH spectrum of a specific relative neutral--cation PAH composition \textit{(1)} small PAHs are being removed (SPR case), and \textit{(2)} large PAHs are being removed (LPR case), in successive steps. A summary of our conclusions follows.

\textit{(i)} The variation in the PAH band intensities as a function of \AvgNc{} has the highest dynamic range  in the SPR case compared to the LPR, suggesting that smaller PAHs have higher impact on the PAH band strengths. This is further supported given the observed variation in the band profiles of the 7.7, 8.6, and 13.5 \micron{} bands in the SPR case, as opposed to the lack of band profile variation in the LPR case. 

\textit{(ii)} Among the individual PAH bands, the 3.3 \micron{} band has the highest dependence on \AvgNc{}, and the largest dynamic range in variation in both the SPR and LPR cases. 

\textit{(iii)} The plateaus in the 5--10 \micron{} and 10--15 \micron{} region show overall declining emission with \AvgNc{} in both the SPR and LPR cases, except for the LPR 10--15 \micron{} plateau which remains almost constant. As with the individual PAH bands, the highest dynamic range in the plateau emissions is seen in the SPR case, suggesting that smaller PAHs are mainly contributing to the plateau emission, especially in the 10--15 \micron{} wavelength region.

\textit{(iv)} The variation in the PAH intensity ratios used as proxies of the PAH size (3.3/(11.2+11.0) \micron{} band ratio) and charge state (6.2/(11.2+11.0), 7.7/(11.2+11.0), and 8.6/(11.2+11.0) \micron{} PAH band ratios) is governed by the feature with the highest variance in a given ratio. The 7.7/(11.0+11.2) ratio presents the least amount of variance with the lowest dynamic range when considering the SPR and LPR cases as a whole, rendering the 7.7/(11.0+11.2) ratio as the better choice for tracing PAH charge.

\textit{(v)} The 3.3/(11.2+11.0) \micron{} PAH band ratio is the only ratio that has both: \textit{(a)} a monotonic variance, and \textit{(b)} fully separated values among the SPR and LPR scenarios. The monotonic variance if driven by the monotonic dependence of the 3.3 \micron{} band with \AvgNc{}. This highlights the efficiency of the 3.3/(11.2+11.0) \micron{} band ratio as PAH size tracer, but also allows the characterization or discrimination of the dominant scenario of processing or formation in a given region or within a source.

\textit{(vi)} The PAH intensity (ratio) and plateau variations as a function of \AvgNc{} is similar among spectra of different ionization fractions (i.e., purely neutral, N75C25, N50C50, N25C75, and purely cationic, spectra) suggesting that the observed trends in SPR and LPR case with in/de-creasing \AvgNc{} is governed by the small/large PAH size cut-off.

\textit{(vii)} The PAH intensity ratio values are dependent on the consideration and removal of a plateau component, with the exception of the 6.2/(11.0+11.2) ratio which shows only minimal variation.

\textit{(viii)} We present new diagnostic PAH charge -- size spaces from the individual and combined SPR and LPR analysis, parameterized on the average (\AvgNc), minimum ($\mathrm{N_{C\,max}}$), and maximum ($\mathrm{N_{C\,min}}$) PAH size of a PAH size distribution, which can provide insights on the average, maximum, or minimum \Nc{} within a an astrophysical source.

\section*{Acknowledgements}
We would like to thank the anonymous referee for the constructive comments and suggestions that have improved the clarity of this paper. A.M.'s research was supported by an appointment to the NASA Postdoctoral Program at NASA Ames Research Center, administered by the Oak Ridge Associated Universities through a contract with NASA. E.P. acknowledges support from an NSERC Discovery Grant and a Western Science and Engineering Research Board (SERB) Accelerator Award. A.R. gratefully acknowledges support from the directed Work Package at NASA Ames titled: ``Laboratory Astrophysics -- The NASA Ames PAH IR Spectroscopic Database''.

\section*{Data Availability Statement}
    The analysis products of this work will be shared on a reasonable request to the corresponding author.

\bibliographystyle{mnras}
\bibliography{Bibliography_PAHs}
	
\appendix	

\section{PAHdb sample UIDs} \label{app:UIDs}

Table \ref{tab:UIDsv320} presents the PAHdb UIDs of the molecules used in this work.

\begin{table*}
\caption{The PAHdb v3.20 UIDs for neutral (N) and cation (C) molecules.} \label{tab:UIDsv320}
\begin{tabular}{@{}cccccccccccc}
  \hline
  \multicolumn{1}{|c|}{PAH} & \multicolumn{2}{c}{UIDs} &
  \multicolumn{1}{|c|}{PAH} & \multicolumn{2}{c}{UIDs} &
  \multicolumn{1}{|c|}{PAH} & \multicolumn{2}{c}{UIDs} &
  \multicolumn{1}{|c|}{PAH} & \multicolumn{2}{c}{UIDs} \\
    & N & C & & N & C & & N & C & & N & C \\
  \hline
  C$_{32}$H$_{14}$ & 4 & 5 &    C$_{32}$H$_{18}$ & 535 & 544 &  C$_{87}$H$_{23}$ & 649 & 647 & C$_{35}$H$_{15}$ & 3230 & 3230  \\  
  C$_{48}$H$_{18}$ & 35 & 36 &  C$_{32}$H$_{18}$ & 536 & 546 &  C$_{119}$H$_{27}$ & 653 & 651 & C$_{37}$H$_{15}$ & 3232 & 3234  \\ 
  C$_{54}$H$_{18}$ & 37 & 38 &  C$_{82}$H$_{24}$ & 561 & 562 &  C$_{67}$H$_{21}$ & 656 & 654 &  C$_{43}$H$_{17}$ & 3235 & 3236 \\  
  C$_{27}$H$_{13}$ & 66 & 65 &  C$_{98}$H$_{28}$ & 565 & 566 &  C$_{146}$H$_{30}$ & 752 & 753& C$_{45}$H$_{17}$ & 3238 & 3239  \\
  C$_{47}$H$_{17}$ & 76 & 77 &  C$_{98}$H$_{28}$ & 567 & 568 &  C$_{144}$H$_{30}$ & 756 & 757& C$_{51}$H$_{19}$ & 3241 & 3242  \\
  C$_{59}$H$_{19}$ & 89 & 88 &  C$_{32}$H$_{14}$ & 591 & 592 &  C$_{142}$H$_{30}$ & 760 & 761 &   C$_{55}$H$_{19}$ & 3244 & 3245 \\  
  C$_{48}$H$_{20}$ & 100 & 101 & C$_{66}$H$_{20}$ & 594 & 595 & C$_{142}$H$_{30}$ & 763 & 764 &   C$_{59}$H$_{21}$ & 3247 & 3248 \\ 
  C$_{96}$H$_{24}$ & 108 & 111 & C$_{66}$H$_{20}$ & 597 & 598 & C$_{142}$H$_{30}$ & 766 & 767 &   C$_{65}$H$_{21}$ & 3249 & 3250 \\ 
  C$_{66}$H$_{20}$ & 115 & 117 & C$_{66}$H$_{20}$ & 600 & 601 &  C$_{140}$H$_{30}$ & 769 & 770 &   C$_{67}$H$_{21}$ & 3252 & 3253 \\
  C$_{78}$H$_{22}$ & 120 & 122 & C$_{66}$H$_{20}$ & 603 & 604 &  C$_{138}$H$_{30}$ & 772 & 773 &   C$_{67}$H$_{23}$ & 3255 & 3256 \\
  C$_{36}$H$_{16}$ & 128 & 129 & C$_{112}$H$_{26}$ & 606 & 607 & C$_{26}$H$_{16}$ & 814 & 815 &    C$_{73}$H$_{21}$ & 3257 & 3258 \\
  C$_{40}$H$_{18}$ & 131 & 132 & C$_{150}$H$_{30}$ & 612 & 613 & C$_{30}$H$_{18}$ & 817 & 818 &    C$_{77}$H$_{23}$ & 3260 & 3261 \\
  C$_{42}$H$_{22}$ & 137 & 138 & C$_{216}$H$_{36}$ & 615 & 621 & C$_{34}$H$_{20}$ & 820 & 821 &   C$_{79}$H$_{23}$ & 3263 & 3264 \\ 
  C$_{44}$H$_{20}$ & 140 & 141 & C$_{170}$H$_{32}$ & 619 & 623 & C$_{34}$H$_{16}$ & 3161 & 3162 & C$_{83}$H$_{23}$ & 3265 & 3266 \\ 
  C$_{48}$H$_{22}$ & 143 & 144 & C$_{40}$H$_{16}$ & 625 & 626 &  C$_{38}$H$_{16}$ & 3164 & 3165 & C$_{85}$H$_{23}$ & 3268 & 3269 \\ 
  C$_{48}$H$_{20}$ & 146 & 147 & C$_{128}$H$_{28}$ & 631 & 632 & C$_{46}$H$_{18}$ & 3169 & 3170 & C$_{87}$H$_{25}$ & 3271 & 3272 \\ 
  C$_{30}$H$_{14}$ & 152 & 153 & C$_{71}$H$_{21}$ & 659 & 657 &  C$_{52}$H$_{18}$ & 3173 & 3174 & C$_{91}$H$_{25}$ & 3273 & 3274 \\ 
  C$_{36}$H$_{16}$ & 154 & 155 & C$_{96}$H$_{23}$ & 693 & 694 &  C$_{54}$H$_{20}$ & 3176 & 3177 & C$_{93}$H$_{25}$ & 3275 & 3276 \\ 
  C$_{42}$H$_{22}$ & 156 & 158 & C$_{96}$H$_{23}$ & 696 & 697 &  C$_{56}$H$_{20}$ & 3179 & 3180 &   C$_{95}$H$_{25}$ & 3277 & 3278\\
  C$_{110}$H$_{26}$ & 162 & 163 & C$_{96}$H$_{22}$ & 699 & 700 & C$_{58}$H$_{20}$ & 3182 & 3183 &   C$_{95}$H$_{27}$ & 3280 & 3281\\
  C$_{112}$H$_{26}$ & 165 & 166 & C$_{66}$H$_{18}$ & 714 & 716 & C$_{62}$H$_{20}$ & 3185 & 3186 &   C$_{99}$H$_{25}$ & 3282 & 3283\\
  C$_{130}$H$_{28}$ & 168 & 169 & C$_{45}$H$_{15}$ & 718 & 721 & C$_{64}$H$_{20}$ & 3188 & 3189 &    C$_{103}$H$_{27}$ & 3285 & 3286\\
  C$_{102}$H$_{26}$ & 177 & 178 & C$_{80}$H$_{20}$ & 719 & 722 &  C$_{64}$H$_{22}$ & 3191 & 3192 &   C$_{107}$H$_{27}$ & 3287 & 3288\\
  C$_{102}$H$_{26}$ & 180 & 181 &  C$_{63}$H$_{21}$ & 727 & 730 & C$_{72}$H$_{22}$ & 3194 & 3195 &   C$_{107}$H$_{27}$ & 3289 & 3290\\
  C$_{110}$H$_{30}$ & 183 & 184 &  C$_{112}$H$_{28}$ & 728 & 731 & C$_{76}$H$_{22}$ & 3197 & 3198 &    C$_{111}$H$_{27}$ & 3291 & 3292\\
  C$_{120}$H$_{36}$ & 186 & 187 &  C$_{210}$H$_{36}$ & 742 & 740 & C$_{80}$H$_{22}$ & 3200 & 3201 &   C$_{113}$H$_{27}$ & 3293 & 3294\\
  C$_{24}$H$_{14}$ & 204 & 205 &   C$_{148}$H$_{30}$ & 743 & 744 & C$_{82}$H$_{24}$ & 3203 & 3204 & C$_{115}$H$_{29}$ & 3296 & 3297\\
  C$_{24}$H$_{14}$ & 206 & 207 &  C$_{146}$H$_{30}$ & 746 & 747 &  C$_{88}$H$_{24}$ & 3206 & 3207 & C$_{115}$H$_{27}$ & 3298 & 3299\\
  C$_{24}$H$_{14}$ & 208 & 209 &   C$_{146}$H$_{30}$ & 749 & 750 & C$_{94}$H$_{24}$ & 3211 & 3212 & C$_{119}$H$_{29}$ & 3300 & 3301\\
  C$_{22}$H$_{14}$ & 301 & 302 &  C$_{190}$H$_{34}$ & 634 & 738 &  C$_{102}$H$_{26}$ & 3214 & 3215 & C$_{121}$H$_{29}$ & 3302 & 3303\\
  C$_{22}$H$_{14}$ & 305 & 306 &  C$_{48}$H$_{18}$ & 635 & 636 &   C$_{108}$H$_{26}$ & 3217 & 3218 & C$_{127}$H$_{31}$ & 3304 & 3305\\
  C$_{22}$H$_{14}$ & 307 & 308 &  C$_{90}$H$_{24}$ & 638 & 639 &   C$_{120}$H$_{28}$ & 3220 & 3221 & & & \\
  C$_{40}$H$_{22}$ & 533 & 540 &  C$_{144}$H$_{30}$ & 641 & 642 &  C$_{148}$H$_{30}$ & 3223 & 3224 & & & \\
  C$_{36}$H$_{20}$ & 534 & 542 &  C$_{57}$H$_{19}$ & 646 & 644 &   C$_{31}$H$_{15}$ & 3226 & 3227 & & & \\
  
\hline
\end{tabular}                                                
\end{table*}

\section{Plateau subtracted spectra}

\begin{figure*}
		\begin{center}
			\hspace*{-0.4cm}\includegraphics*[keepaspectratio=true,scale=0.45]{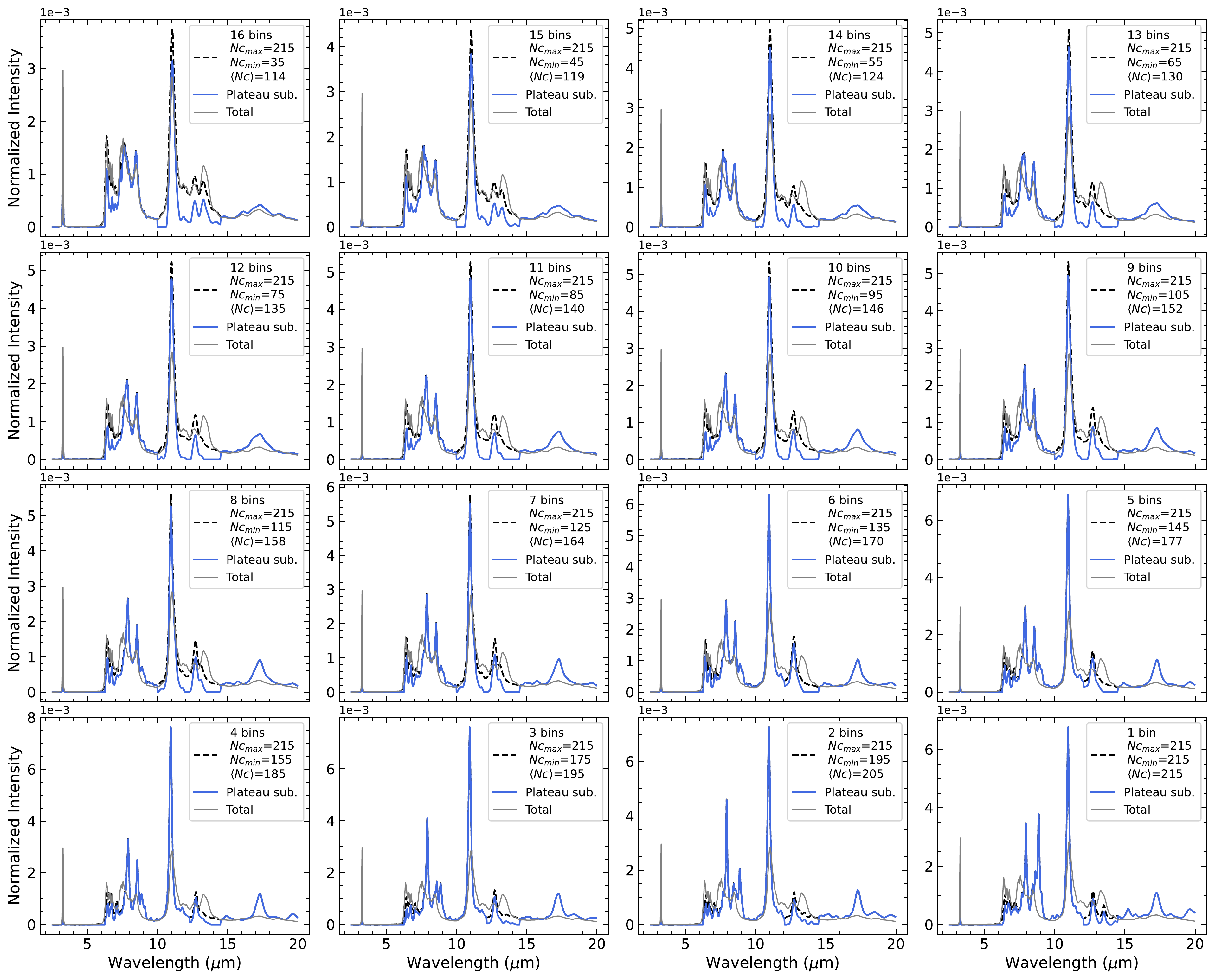}
			\caption{The plateau-subtracted spectra for the SPR N50C50 case (blue color), as presented in Figure \ref{fig:SPR_binspec}. The non-subtracted spectrum in each panel is shown with a dashed black line, and the total spectrum with a gray line.}
			\label{fig:SPR_plr}
		\end{center}
	\end{figure*}

\begin{figure*}
		\begin{center}
			\hspace*{-0.4cm}\includegraphics*[keepaspectratio=true,scale=0.45]{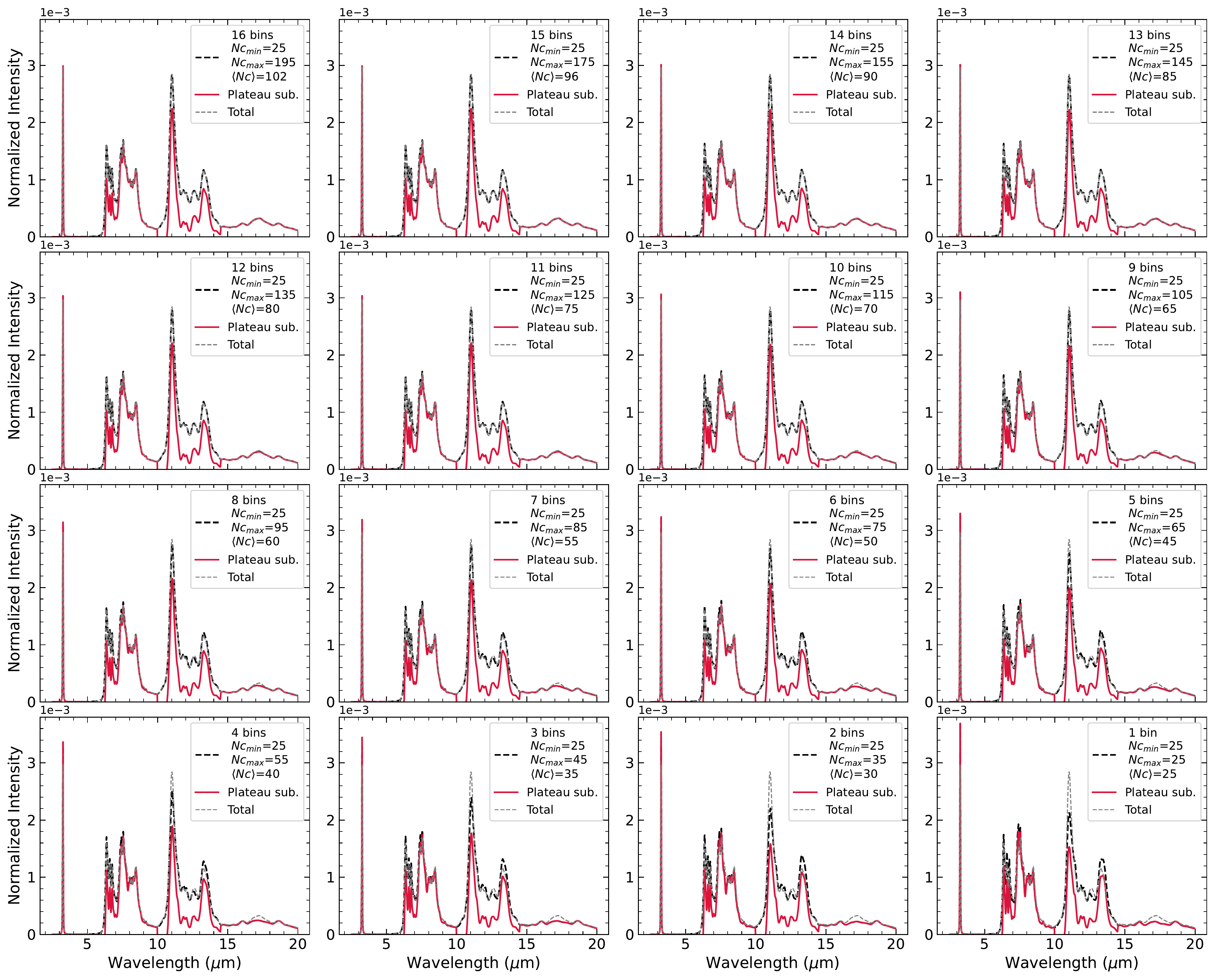}
			\caption{The plateau-subtracted spectra for the LPR N50C50 case (red color), as presented in Figure \ref{fig:LPR_binspec}. The non-subtracted spectrum in each panel is shown with a dashed black line, and the total spectrum with a gray line.}
			\label{fig:LPR_plr}
		\end{center}
	\end{figure*}

Figures \ref{fig:SPR_plr} and \ref{fig:LPR_plr} present the plateau subtracted spectra in each step of the SPR and LPR N50C50 respectively, following Figures \ref{fig:SPR_binspec} and \ref{fig:LPR_binspec}, and the methodology in Section \ref{sec:EmissionLines}.

\section{Variation in the 7.7 \micron{} PAH band} \label{app:77_region}

\begin{figure*}
		\begin{center}
			\hspace*{-0.4cm}\includegraphics*[keepaspectratio=true,scale=0.45]{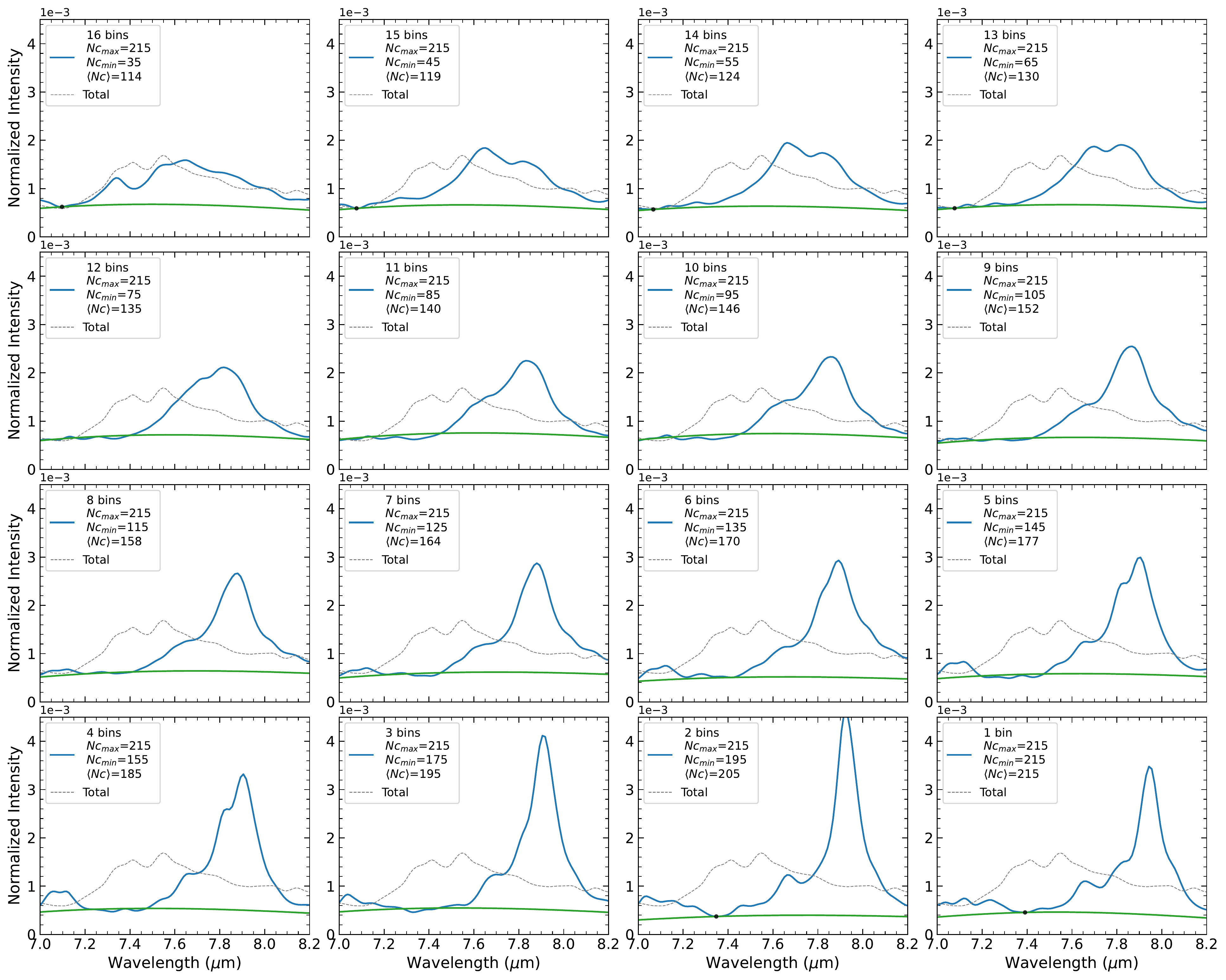}
			\caption{The variation in the 7.7 \micron{} PAH band in the different steps of the SPR N50C50 case spectra (blue line), as presented in Figure \ref{fig:SPR_binspec}. The total synthesized spectrum is shown with a dashed gray line in all panels, and the green line is the local continuum in the 5-10 \micron{} range defined with the spline method.}
			\label{fig:SPR_77}
		\end{center}
	\end{figure*}

\begin{figure*}
		\begin{center}
			\hspace*{-0.4cm}\includegraphics*[keepaspectratio=true,scale=0.45]{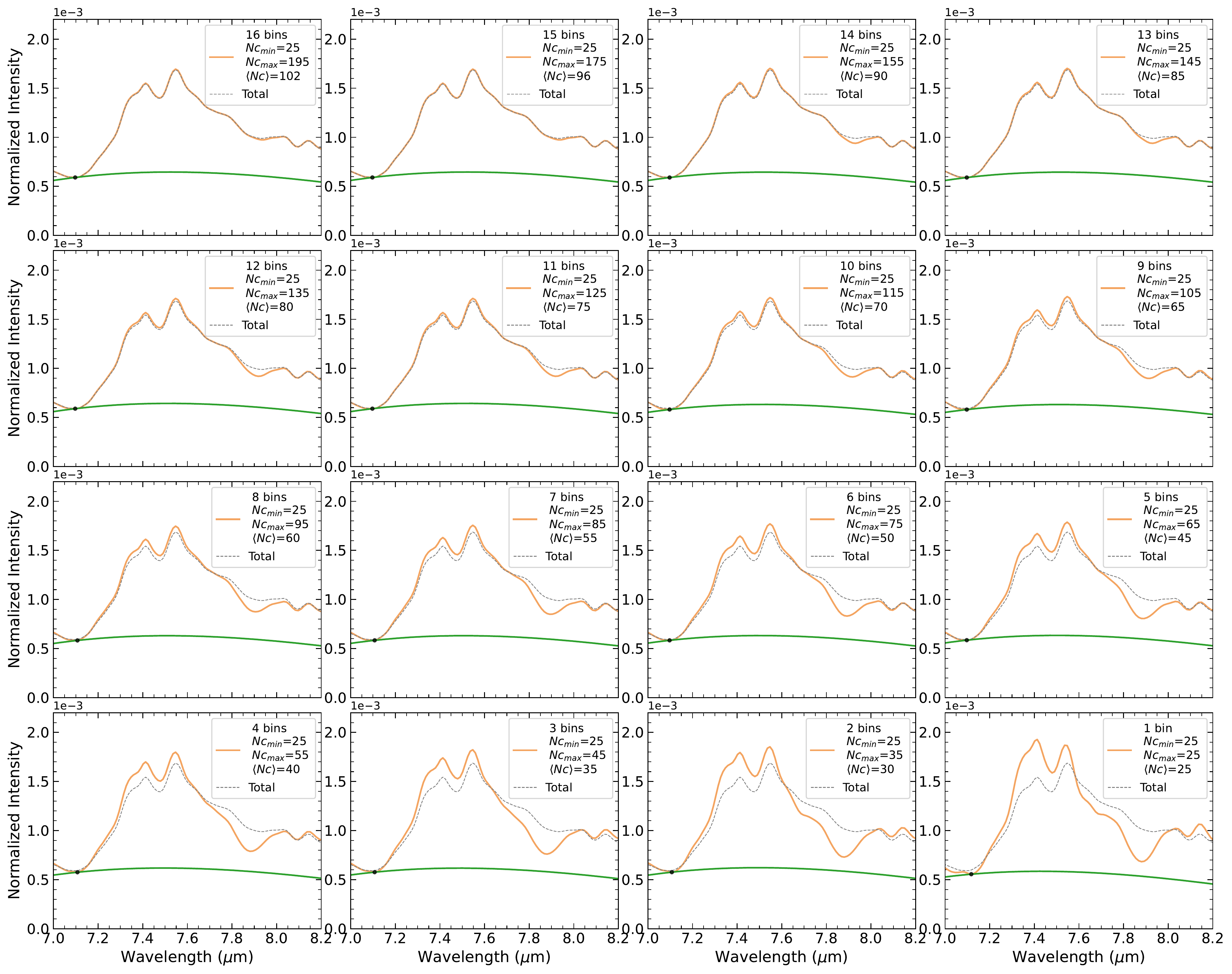}
			\caption{The variation in the 7.7 \micron{} PAH band in the different steps of the LPR N50C50 case spectra (orange line), as presented in Figure \ref{fig:LPR_binspec}. The total synthesized spectrum is shown with a dashed gray line in all panels, and the green line is the local continuum in the 5-10 \micron{} range defined with the spline method.}
			\label{fig:LPR_77}
		\end{center}
	\end{figure*}

The profile variation and position shift of the 7.7. \micron{} PAH band in the N50C50 SPR and LPR cases is presented in Figures \ref{fig:SPR_77} and \ref{fig:LPR_77}.

\section{PAH intensity variations for different ionization fractions} \label{sec:Int_AvgNc_various}

\begin{figure*}
		\begin{center}
			\hspace*{-0.1cm}\includegraphics[keepaspectratio=true,scale=0.6]{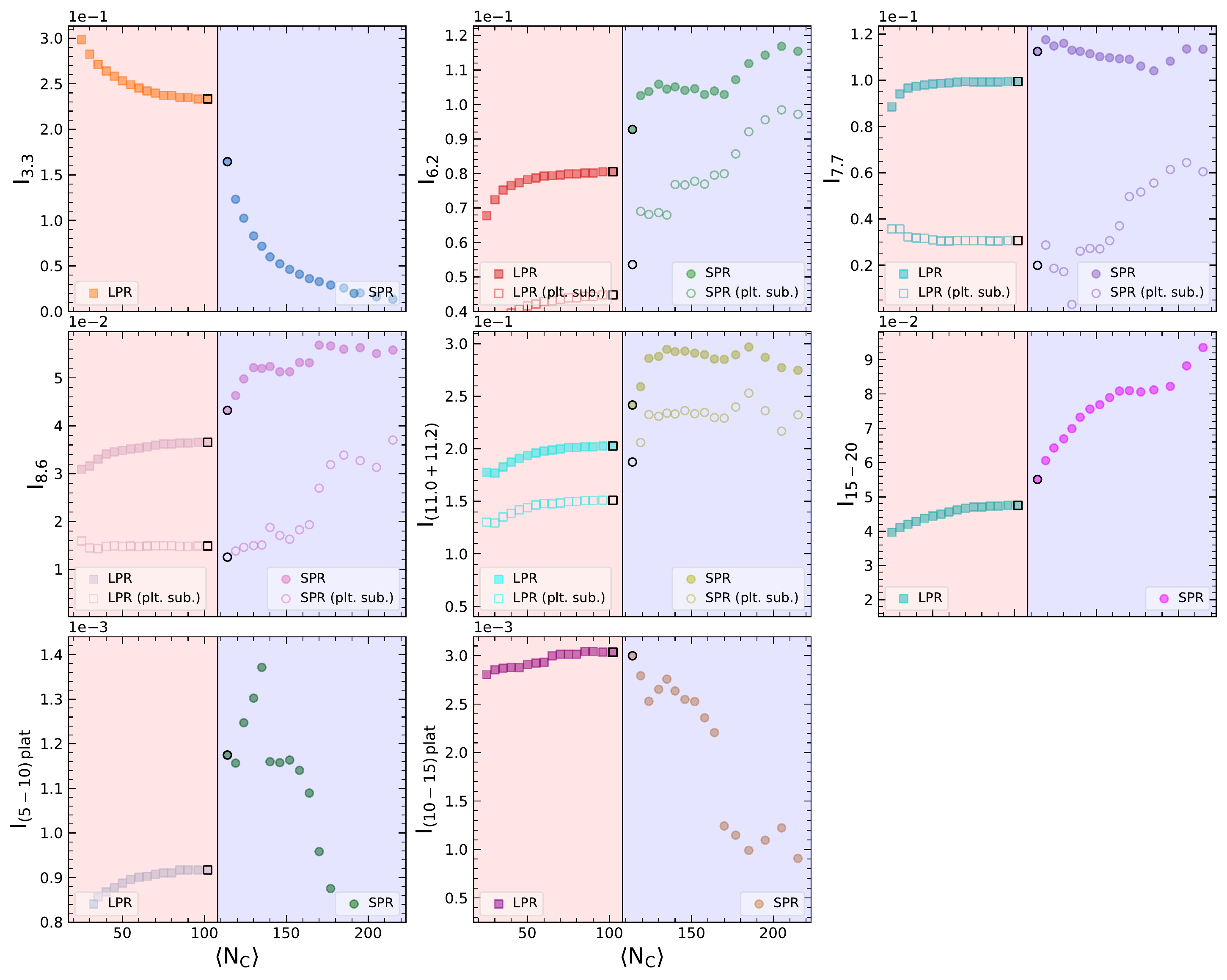}
			\caption{Intensity variations of PAH features and plateaus normalized to the total flux of each spectrum for the Neutral case, as a function of \AvgNc, for the small PAHs removed (SPR) and large PAHs removed (LPR) cases. Open circles and squares, where present, indicate plateau subtracted fluxes for the SPR and LPR cases respectively. Black edged circles and squares indicate the initial resulting spectrum after the removal of the first bin's spectrum in each case (SPR and LPR) from the total spectrum. All intensities are normalized to the total flux of each individual spectrum.}
			\label{fig:Int_AvgNc_Neutral}
		\end{center}
	\end{figure*}

\begin{figure*}
		\begin{center}
			\hspace*{-0.1cm}\includegraphics[keepaspectratio=true,scale=0.6]{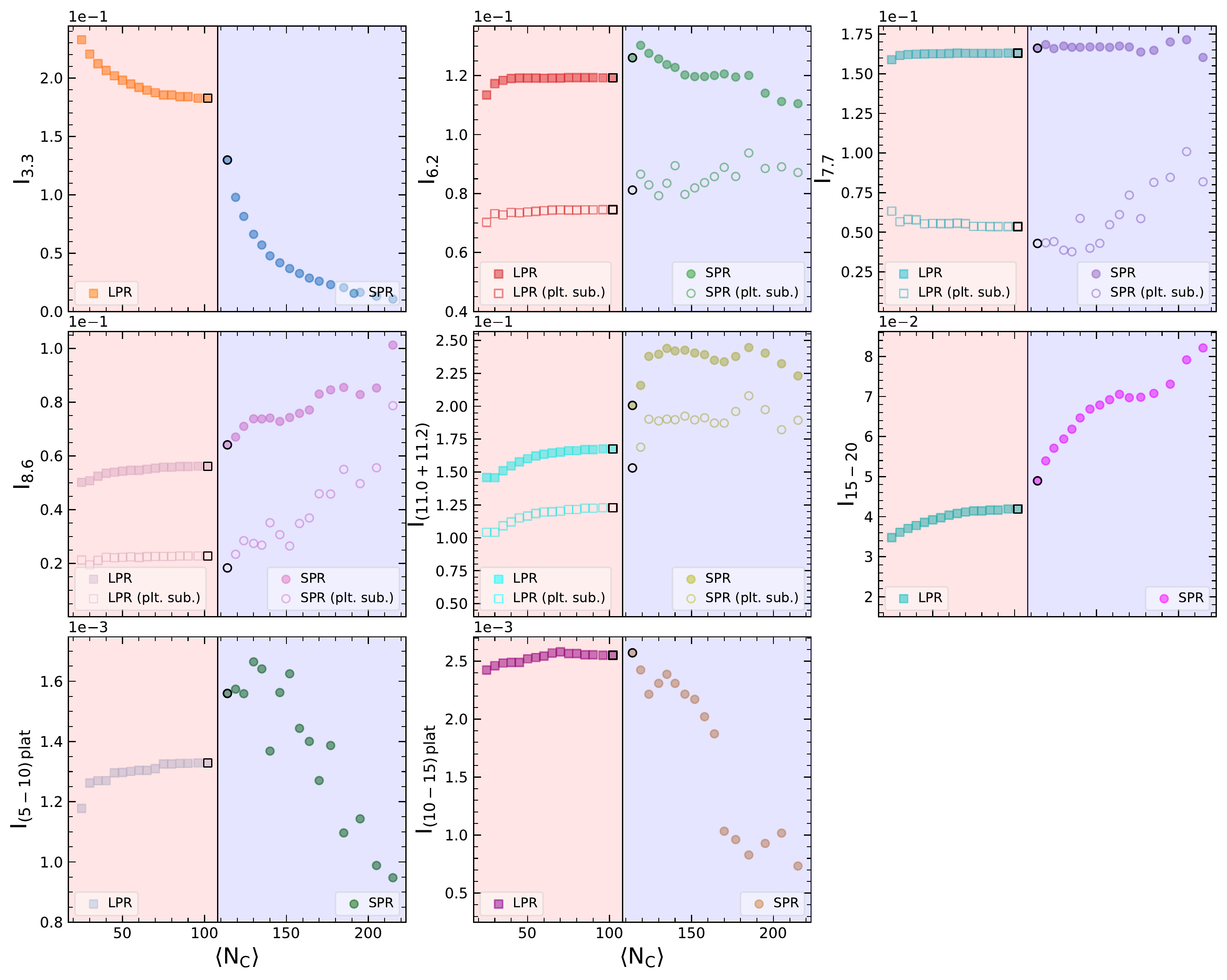}
			\caption{Intensity variations of PAH features and plateaus normalized to the total flux of each spectrum for the N75C25 case, as a function of \AvgNc, for the small PAHs removed (SPR) and large PAHs removed (LPR) cases. Open circles and squares, where present, indicate plateau subtracted fluxes for the SPR and LPR cases respectively. Black edged circles and squares indicate the initial resulting spectrum after the removal of the first bin's spectrum in each case (SPR and LPR) from the total spectrum. All intensities are normalized to the total flux of each individual spectrum.}
			\label{fig:Int_AvgNc_N75C25}
		\end{center}
	\end{figure*}

\begin{figure*}
		\begin{center}
			\hspace*{-0.1cm}\includegraphics[keepaspectratio=true,scale=0.6]{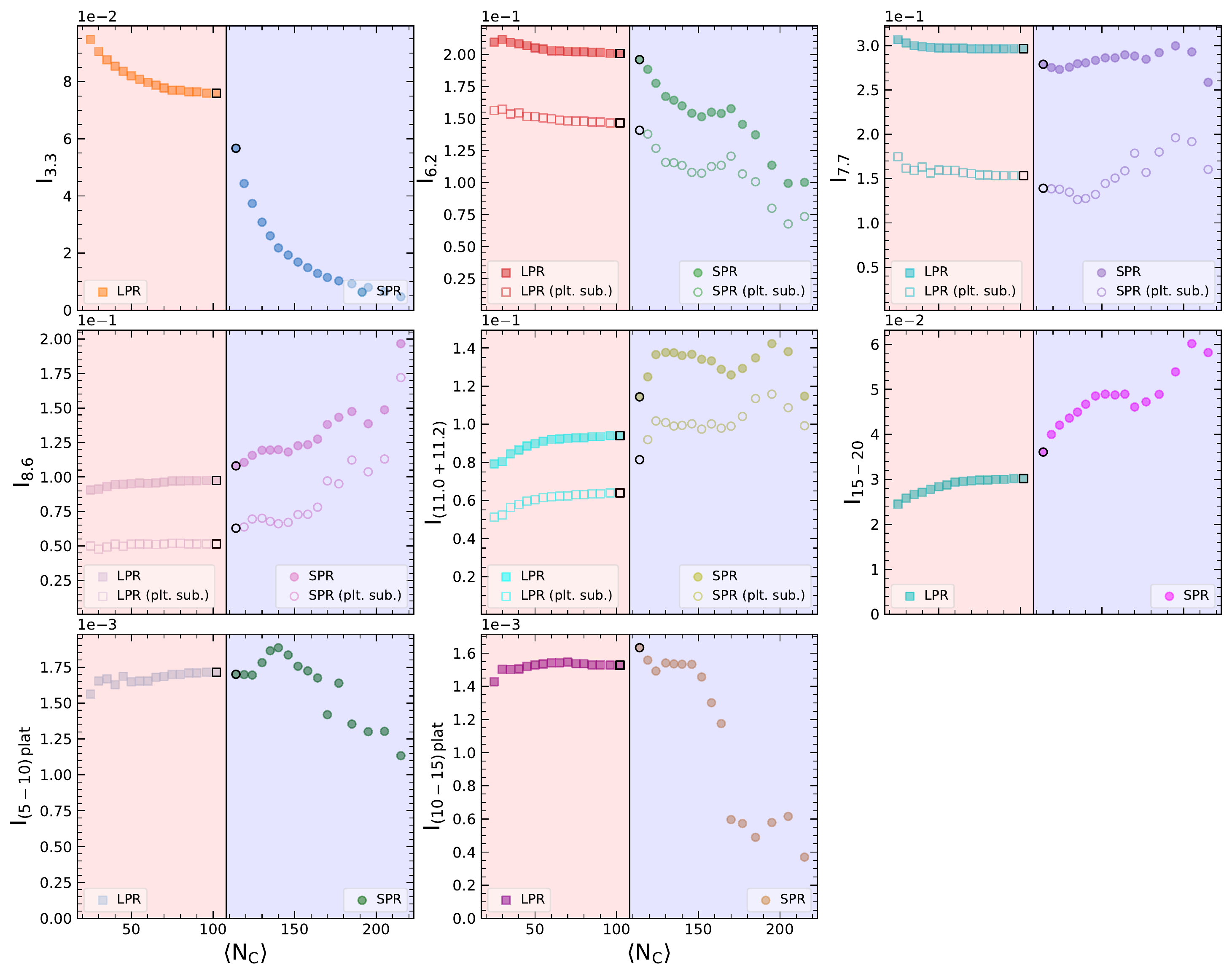}
			\caption{Intensity variations of PAH features and plateaus normalized to the total flux of each spectrum for the N25C75 case, as a function of \AvgNc, for the small PAHs removed (SPR) and large PAHs removed (LPR) cases. Open circles and squares, where present, indicate plateau subtracted fluxes for the SPR and LPR cases respectively. Black edged circles and squares indicate the initial resulting spectrum after the removal of the first bin's spectrum in each case (SPR and LPR) from the total spectrum. All intensities are normalized to the total flux of each individual spectrum.}
			\label{fig:Int_AvgNc_N25C75}
		\end{center}
	\end{figure*}

\begin{figure*}
		\begin{center}
			\hspace*{-0.1cm}\includegraphics[keepaspectratio=true,scale=0.6]{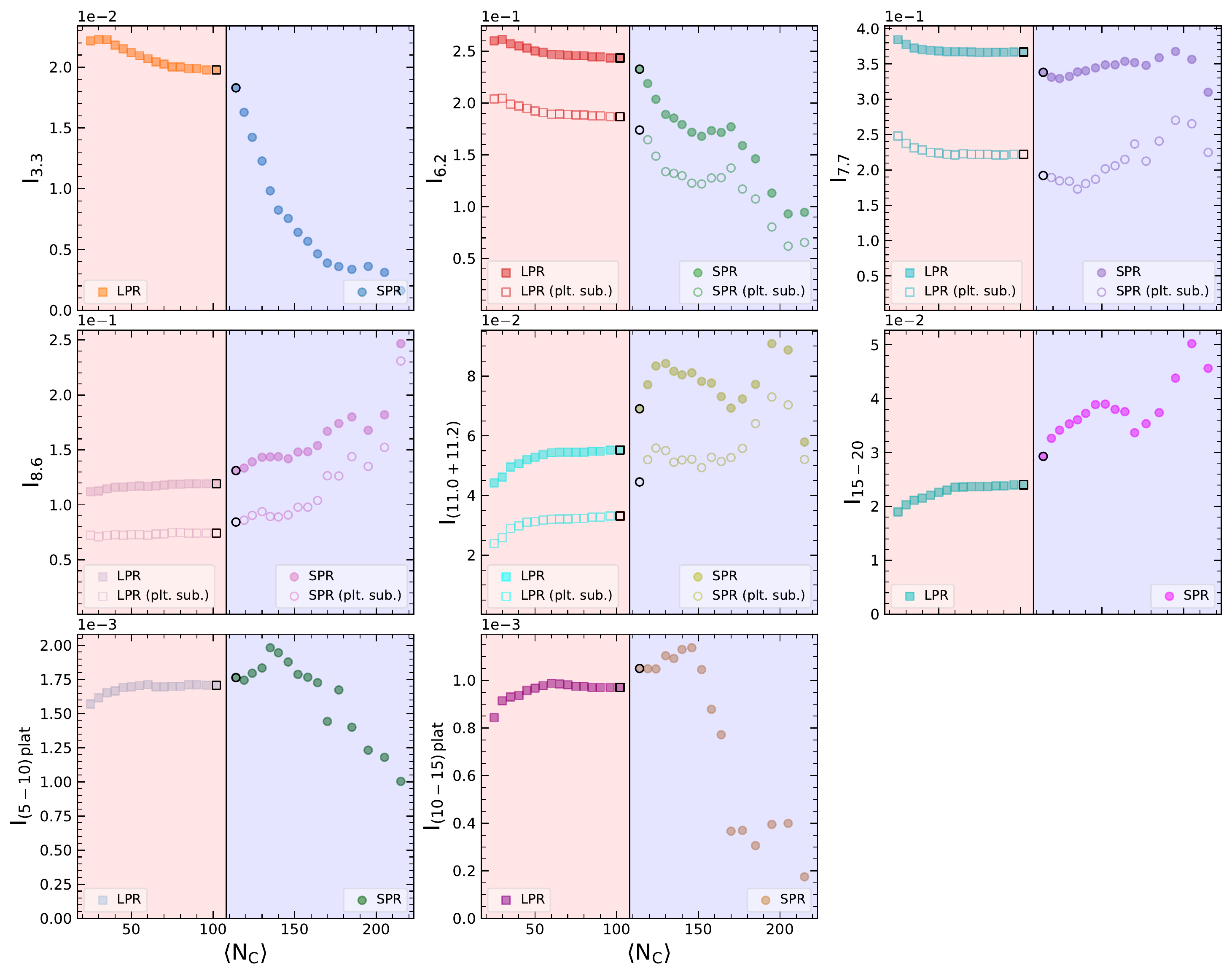}
			\caption{Intensity variations of PAH features and plateaus normalized to the total flux of each spectrum for the Cation case, as a function of \AvgNc, for the small PAHs removed (SPR) and large PAHs removed (LPR) cases. Open circles and squares, where present, indicate plateau subtracted fluxes for the SPR and LPR cases respectively. Black edged circles and squares indicate the initial resulting spectrum after the removal of the first bin's spectrum in each case (SPR and LPR) from the total spectrum. All intensities are normalized to the total flux of each individual spectrum.}
			\label{fig:Int_AvgNc_Cation}
		\end{center}
	\end{figure*}

Figures \ref{fig:Int_AvgNc_Neutral}--\ref{fig:Int_AvgNc_Cation} present the intensity variations of the main PAH features (with and without plateau subtraction), as a function of \AvgNc, for each step of the SPR and LPR cases, for spectra of different PAH ionization fractions, namely purely neutral, N75C25, N25C75, and purely cationic spectra (Section \ref{sec:Spectra}). 

\end{document}